\documentclass[11pt,onecolumn]{article}
\usepackage[letterpaper,margin=0.75in]{geometry}
\usepackage{amsmath,amssymb,amsfonts}
\usepackage{graphicx}
\usepackage{booktabs}
\usepackage{multirow}
\usepackage{array}
\usepackage{cite}
\usepackage[hidelinks]{hyperref}
\usepackage{caption}
\usepackage{subcaption}
\usepackage{xcolor}
\usepackage{url}
\usepackage{setspace}
\usepackage{amsmath}
\usepackage{color}
\usepackage{graphicx} 
\usepackage{caption} 
\usepackage{enumitem}
\usepackage{subcaption}
\usepackage{adjustbox}
\usepackage{amsfonts} 
\usepackage{hyperref}
\usepackage{cite}
\usepackage{amssymb}
\usepackage{algorithmic}
\usepackage{textcomp}
\usepackage{multirow}
\usepackage{comment}
\usepackage{threeparttable}

\usepackage[normalem]{ulem}

\title{\textbf{Multiscale Community-Based Fingerprinting of Signed Functional Networks}}

\author{
    Sema Athamnah$^{1}$,
    Selin Aviyente$^{2}$\\
    \small $^{1}$ Department of Biomedical Engineering, Michigan State University, USA.\\
    \small $^{2}$ Department of Electrical and Computer Engineering, Michigan State University, USA.\\
    \small \texttt{athamnah@msu.edu, aviyente@egr.msu.edu}
}
\date{}

\begin{document}
\maketitle
    \newcommand{\matr}[1]{\mathbf{#1}}
    \newcommand\scalemath[2]{\scalebox{#1}{\mbox{\ensuremath{\displaystyle #2}}}} 
\begin{abstract} 
Objective: Recent studies demonstrate that functional connectomes contain subject-specific signatures, or \textit{fingerprints}, that can identify individuals across repeated sessions and tasks. Existing methods mostly rely on edge-level features that are sensitive to noise, difficult to interpret, and limited in their ability to generalize across tasks and datasets. Methods: We propose a multiscale community-based functional connectome fingerprinting framework that characterizes each individual by the mesoscale structure of their functional networks. We introduce a signed multilayer community detection framework that incorporates both correlated and anti-correlated brain activity to identify subject-specific community structures across tasks and sessions. Graph-theoretic metrics are then computed from the resulting joint community structures to derive low-dimensional community-level fingerprint representations.
Results: The proposed framework is evaluated on 810 healthy control subjects from the Human Connectome Project (HCP). The results show that community-based fingerprints provide a reliable and interpretable substrate for individualized brain characterization across sessions and tasks.
Conclusion: Mesoscale community structure provides meaningful and discriminative subject-specific fingerprints. 
Significance: The proposed framework offers a promising foundation for precision neuroimaging and personalized neuroscience applications.
\end{abstract}


\begin{table}[h!]
    \begin{tabular}{ll}
    \hline
    Notations & Explanations\\
    \hline
    $\mathcal{G}^{\pm}$ & Signed graph. \\
    $N$ & Number of nodes.\\
    $V$ & Set of nodes.\\
    $E$ & Set of edges.\\
    $\matr{A}^{\pm}$ & Positive (negative) adjacency matrix. \\ 
    $\matr{P}^{\pm}$ & Positive (negative) null-model adjacency matrix.\\
    $\mathcal{K}_{i}^{\pm}$ &  Positive (negative) strength of node $i$.\\
    $2\mathcal{T}^{\pm}$ & Total positive (negative) strength of the signed graph.\\
    $S$ & Number of layers.\\
    $\gamma$ & Resolution parameter controlling community size. \\
    $\omega_{jsr}$ & Inter-layer coupling parameter between layer $s$ and $r$ for node $j$. \\
    $\mathbf{g}$ & Community assignment vector of a single-layer network, $\mathbf{g} \in \{1,\ldots,K\}^{N \times 1}$, where $K$ is the number \\&of communities and each element $g_i$ denotes the community label assigned to node $i$.\\
    $\matr{C}$ & Joint community assignment matrix for a multilayer network, $\matr{C} \in \{1,\ldots,K\}^{N \times S}$, where $C_{is}$ is the \\&community label of node $i$ in layer $s$.\\
    $\mathcal{P}$& Set of all possible partitions of $N$ nodes, $\mathcal{P}=\{P_{1},\ldots, P_{M}\}$, where $M$ is the total number of \\&partitions.\\
    $K_m$& Number of communities in partition $P_m$. \\
    $c_k$ & The $k$-th community in partition $P_m$.\\
    $|c_k|$& Number of nodes in community $c_k$.\\
    $p_{is}^{+}(c_k)$ & Ratio of the positive edge strength from node $i$ to nodes in community $c_k$, to the total positive edge \\&strength of node $i$ in layer $s$. \\    
    $\alpha_{s}^{f,m}$ & Community feature vector (fingerprint) for subject $s$, extracted from partition $P_{m}$ using feature $f$, \\&where $\alpha_{s}^{f,m}\in\mathbb{R}^{K_m\times1}$.\\
    $\phi^{f,m}$ & Mean pairwise Euclidean distance between fingerprints extracted using feature $f$ from partition $P_m$. \\
    $\{\alpha_{s}^{f,l}\}_{s=1}^{S}$&Training CBFs for feature $f$ and partition $P_l$.\\
    $\{\tilde{\alpha}_{s}^{f,l}\}_{s=1}^{S}$& Testing CBFs for feature $f$ and partition $P_l$.\\
    \hline 
    \end{tabular}
    \caption{Notation definitions.}
    \label{tab:notations}
\end{table}

\begin{figure*}[!ht]
\centering\includegraphics[width=\textwidth]{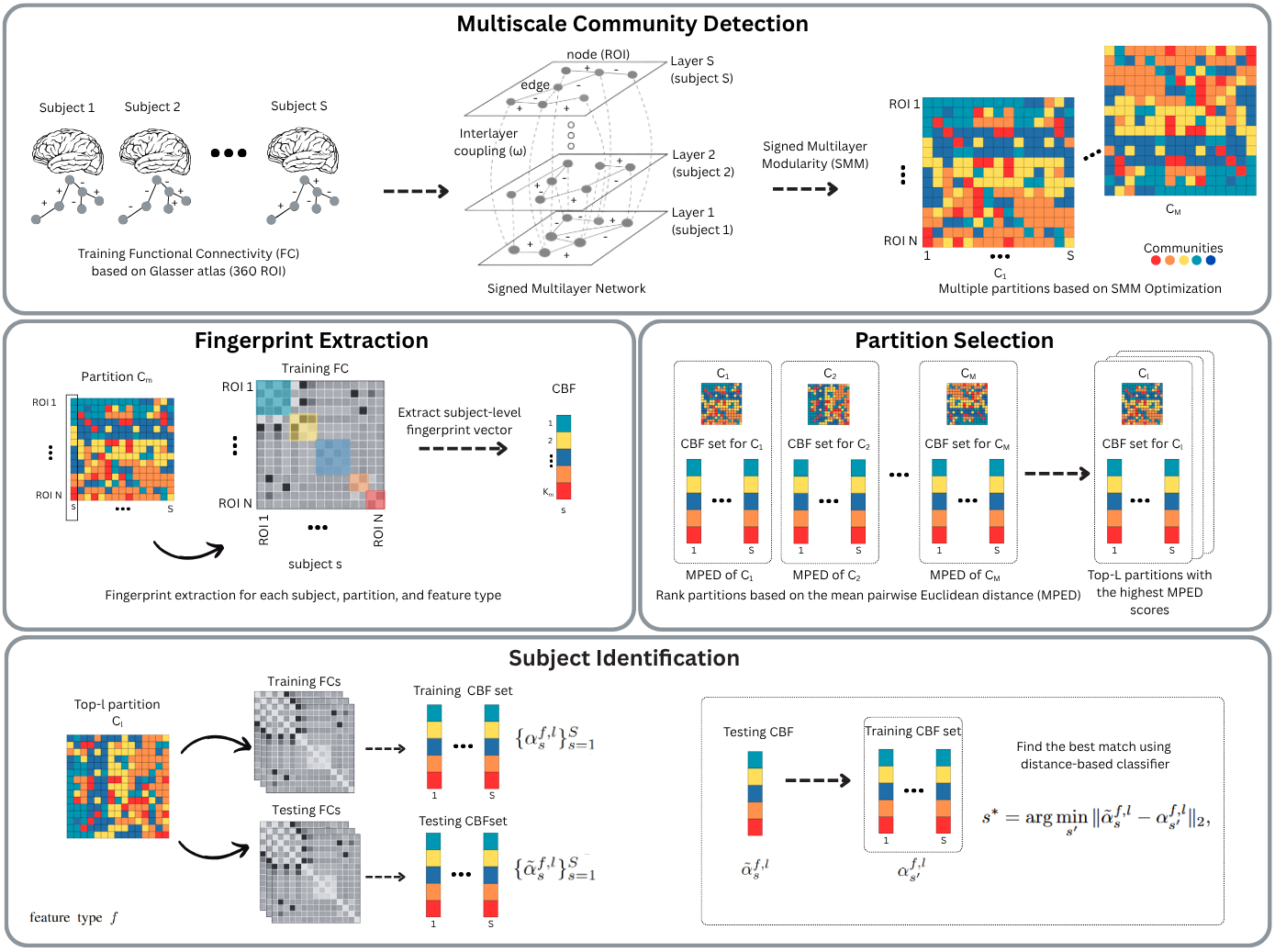} 
\caption{Overview of the proposed community-based fingerprinting framework.
(a) \textbf{Multiscale Community Detection:} A signed multilayer modularity (SMM) optimization framework is applied to generate a set of multiscale partitions by varying the resolution ($\gamma$) and interlayer coupling ($\omega$) parameters (Section~\ref{CD_SMM}).
(b) \textbf{Fingerprint Extraction:} For each partition, community-level graph-theoretic features are extracted from each subject's functional connectivity (FC) network to construct community-based fingerprint (CBF) vectors (Section~\ref{CBF_ext}).
(c) \textbf{Partition Selection:} The resulting partitions are ranked according to the mean pairwise Euclidean distance (MPED) of their corresponding CBF vectors, and the top-$L$ partitions are selected (Section~\ref{Partition Selection}).
(d) \textbf{Subject Identification:} The selected top-$L$ partitions are used to construct the training and testing CBFs from the corresponding training and testing FC networks. For each selected partition and feature type, each testing CBF vector is matched to its corresponding training CBF vector using a distance-based nearest-neighbor classifier for subject identification (Section~\ref{Sub_iden}).}
    \label{fig:overview}
\end{figure*}

\section{Introduction}
\label{sec:intro}
Understanding individual variability in brain function is a central goal in cognitive and clinical neuroscience. Functional connectivity (FC), derived from pairwise correlation of fMRI time series across brain regions, provides a powerful representation of large-scale neural communication \cite{biswal1995functional}.  Recent research has shown that FC patterns are unique to the individual and are stable across sessions and tasks, providing a useful representation, known as functional connectome fingerprint, for subject identification \cite{finn2015functional, gratton2018functional}. These subject-specific features of the connectome have opened new opportunities for linking FC patterns to cognitive traits and behavioral phenotypes \cite{finn2021beyond, amico2018quest}.

Early fingerprinting approaches rely on whole FC matrices or selected specific subnetworks for fingerprinting \cite{finn2015functional,pena2018spatiotemporal}. These methods calculate the correlation between  a target FC matrix and  each reference FC matrix to determine the best match. 

On the other hand, feature extraction-based methods learn low-dimensional embeddings such as principal component analysis (PCA)-based features \cite{amico2018quest}, graph embedding-based methods \cite{abbas2020geff}, and tensor-based features \cite{carvalho2025functional}. In addition, machine learning-based approaches have been widely explored for functional connectome fingerprinting \cite{hannum2023high,cai2021functional,griffa2022brain,lee2024discovering,lu2024brain}.
A key challenge in advancing connectome fingerprinting is that most existing methods rely on univariate or edge-level features, rather than multivariate features that capture functionally coherent network organization \cite{amico2018quest}. This is problematic since individual differences in connectivity are known to arise from coordinated interactions within and between large-scale brain systems rather than from isolated edges. For example, the default mode network (DMN), frontoparietal control network (FPN), and dorsal attention networks (DAN) consistently contribute to individual distinctiveness in resting and task states \cite{st2023functional}. Capturing such structured patterns of variability is essential for both accurate identification and meaningful clinical interpretation.
Thus, there is a need for a fingerprinting framework that identifies individualized, functionally meaningful subgraphs rather than isolated connections while also providing interpretability. In this paper, we address these challenges by developing a community-based functional connectome fingerprinting approach that integrates signed multilayer community detection with subject-level graph-theoretic feature extraction \cite{sporns2016modular,Mucha2010,betzel2017multi}. This framework shifts identifiability from edges to multivariate community-level topological signatures, enabling more robust and interpretable characterization of individual differences in brain organization.

The major contributions of this paper are three-fold. First, we formulate fingerprinting at the mesoscale by representing each subject through data-driven multivariate descriptors of functional communities rather than univariate edge weights, making subnetworks the fundamental unit of analysis and providing a more interpretable, neurobiologically meaningful representation of individual variability (Section~\ref{sec:method}) \cite{amico2018quest,abbas2020geff}.
Second, we introduce a signed multilayer modularity formulation that explicitly incorporates both positively correlated and anti-correlated interactions, in contrast to traditional multilayer modularity approaches \cite{Mucha2010}, which only focus on positive connections (Section ~\ref{CD_SMM}). This formulation provides a more robust characterization of community structure and its variability across individuals.
Finally, we introduce the Discriminative Connectivity Map (DCM), which localizes the connectivity patterns contributing most strongly to subject identification and renders them at either the ROI or the functional-system level for a single subject or a group (Section~\ref{C-ICC}). DCMs provide the interpretive link between the identifiability statistics and known functional neuroanatomy ~\cite{amico2018quest,abbas2020geff}.
Together, these contributions result in a more interpretable, mechanistically meaningful, and individualized approach to brain fingerprinting, moving the field closer to precision neuroscience.

\section{Background}
\label{sec:Back}
\subsection{Functional Connectivity Networks}
Functional connectivity (FC) quantifies the co-fluctuations between regions within and across brain networks. For BOLD signals, connectivity is typically measured using Pearson’s correlation, yielding a signed graph $\mathcal{G}^{\pm}=(V, E, \matr{A}^{\pm})$, where $V$ is the node set corresponding to brain regions with $|V|=N$, $E$ is the edge set describing functional dependencies, and $\matr{A}^{\pm}\in\mathbb{R}^{N\times N}$ is the weighted, undirected signed adjacency matrix defined as $A_{ij}^{\pm}=A_{ij}^{+}-A_{ij}^{-}$, where $A_{ij}^{+}$ and $A_{ij}^{-}$ denote the $(i,j)$-th entries of the positive and negative adjacency matrices, respectively, with $A_{ij}^{\pm}\in[-1,1]$~\cite{sporns2016modular}. 

\vspace{-0.1in}
\subsection{Signed Modularity Function }
Modularity maximization is one of the most widely used methods for community detection \cite{article,Fortunato2016}. 
In signed networks, communities are defined based on structural balance, where nodes within a community share positive connections while connections between communities are predominantly negative \cite{davis1967clustering}. The signed modularity function has been defined to be consistent with this intuition by separately modeling the contributions of positive and negative correlations \cite{gomez2009analysis}.


Given a signed graph $\mathcal{G}^{\pm}=(V,E,\matr{A}^{\pm})$, the optimal partition $P$, i.e., the community assignment of each node, is obtained by maximizing the signed modularity over the space of all network partitions $\mathcal{P}$.  We restrict ourselves to hard (non-overlapping) partitions, in which each node is assigned to exactly one community denoted by the community assignment vector, $\mathbf{g} \in \{1,\ldots,K\}^{N \times 1}$ where $K$ is the number of communities.
The signed modularity is then defined as:

\begin{equation} 
    \arg\max_{\mathbf{g}} Q^{\pm}(\mathbf{g}|\gamma;\matr{A}^{\pm},\matr{P}^{\pm}):=\sum_{i,j=1}^{N} (A_{ij}^{\pm} -\gamma P_{ij}^{\pm}) \delta(g_i,g_j),
    \label{eq:SM}
\end{equation}
\noindent where $\gamma$ is the resolution parameter controlling the size of the detected communities, $g_{i}$ denotes the community assignment of node $i$, with $\delta(g_{i},g_{j})$ being the Kronecker delta, which equals $1$ if nodes $i$ and $j$ belong to the same community and $0$ otherwise, and $\matr{P}^{\pm}$ denotes the signed adjacency matrix corresponding to the null-model. In this work, we adopt the signed extension of the Newman--Girvan null model \cite{gomez2009analysis}, defined as $P_{ij}^{\pm}=P_{ij}^{+}-P_{ij}^{-}
=\frac{\mathcal{K}_i^{+}\mathcal{K}_j^{+}}{2\mathcal{T}^{+}}
-\frac{\mathcal{K}_i^{-}\mathcal{K}_j^{-}}{2\mathcal{T}^{-}},$ where $\mathcal{K}_{i}^{+}=\sum_j A_{ij}^{+}$ and $\mathcal{K}_{i}^{-}=\sum_j A_{ij}^{-}$ denote the positive and negative strengths of node $i$, respectively, and the total positive and negative strengths are given by $
2\mathcal{T}^{+}=\sum_{ij}A_{ij}^{+}$, and $
2\mathcal{T}^{-}=\sum_{ij}A_{ij}^{-}.$


\vspace{-0.1in}
\subsection{Multilayer Modularity Function}
Let $\mathcal{G}_{s}=(V, E_{s}, \matr{A}_{s})$ for $s \in \{1,\ldots, S\}$ denote multilayer networks with $S$ layers, where $E_{s}$ and $\matr{A}_{s}\in \mathbb{R}^{N\times N}$ are the edge set and the adjacency matrix for layer $s$, respectively.
The multilayer modularity function incorporates an interlayer coupling parameter $\omega$ to control the coupling, i.e., consistency between layers \cite{Mucha2010,betzel2019community} as:

\begin{equation} 
\begin{gathered}
\arg\max_{\matr{C}}
Q(\matr{C}|\gamma,\omega;\matr{A}_s,\matr{P}_s)
:=\sum_{s,r=1}^{S}\sum_{i,j=1}^{N}\big[\left(A_{ijs}-\gamma P_{ijs}\right)\delta_{sr}+\omega_{jsr}\delta_{j}\big]\delta(C_{is},C_{jr}).
\label{eq:MM}
\end{gathered}\end{equation}

\noindent where $\omega_{jsr}$ is the inter-layer coupling parameter between layers $s$ and $r$ for node $j$, and $\matr{P}_s$ is the Newman-Girvan null model for layer $s$ defined as $P_{ijs}=\frac{\mathcal{K}_{is} \mathcal{K}_{js}}{2 \tau_{s}}$.
We restrict our attention to multiplex networks, in which all layers share the same set of nodes but differ in their topological structure. $\matr{C} \in \{1,\ldots,K\}^{N \times S}$ is the community assignment matrix
where $C_{ij}$ denotes the community label assigned to node $i$ in layer $j$.

\begin{figure*}[!ht]
\centering
\includegraphics[width=\columnwidth]{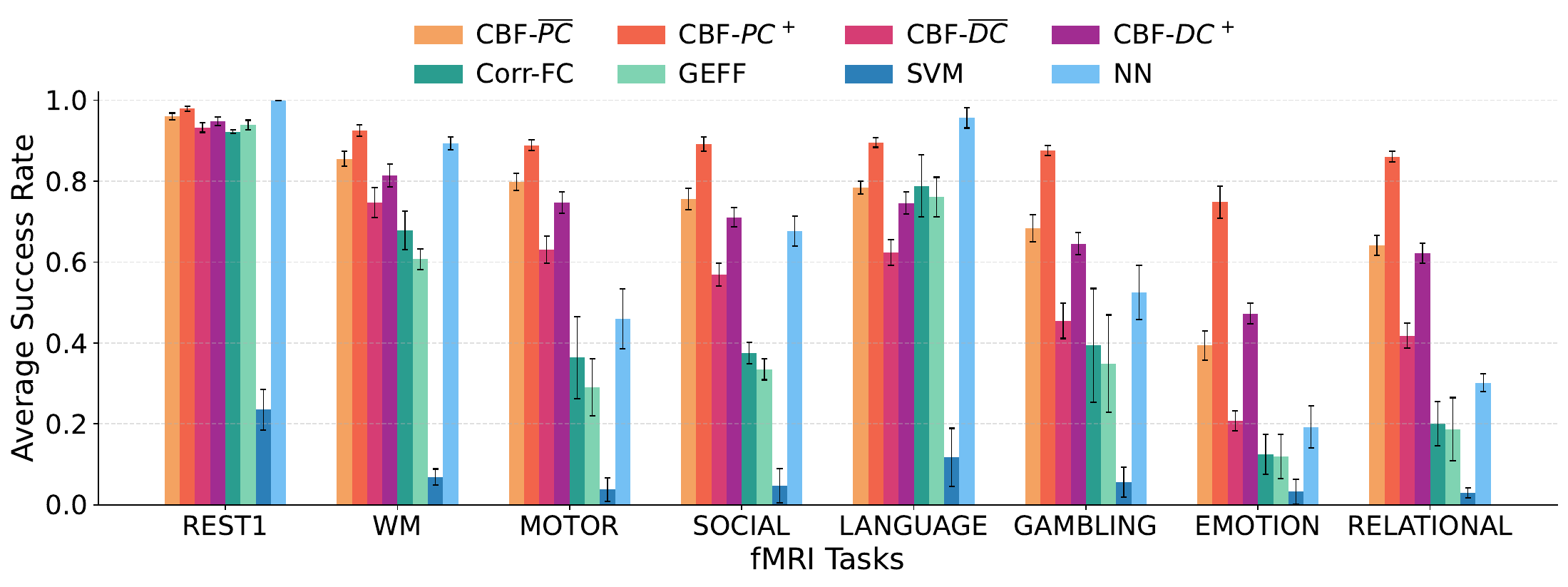}
\caption{Within-task fingerprinting results for a sample size of $S=100$, averaged across 10 randomly selected subject groups. 
}
\label{fig:within_100}
\end{figure*}

\begin{figure*}[!ht]
\centering
\begin{subfigure}[t]{0.49\textwidth}
    \centering
    \includegraphics[width=\linewidth]{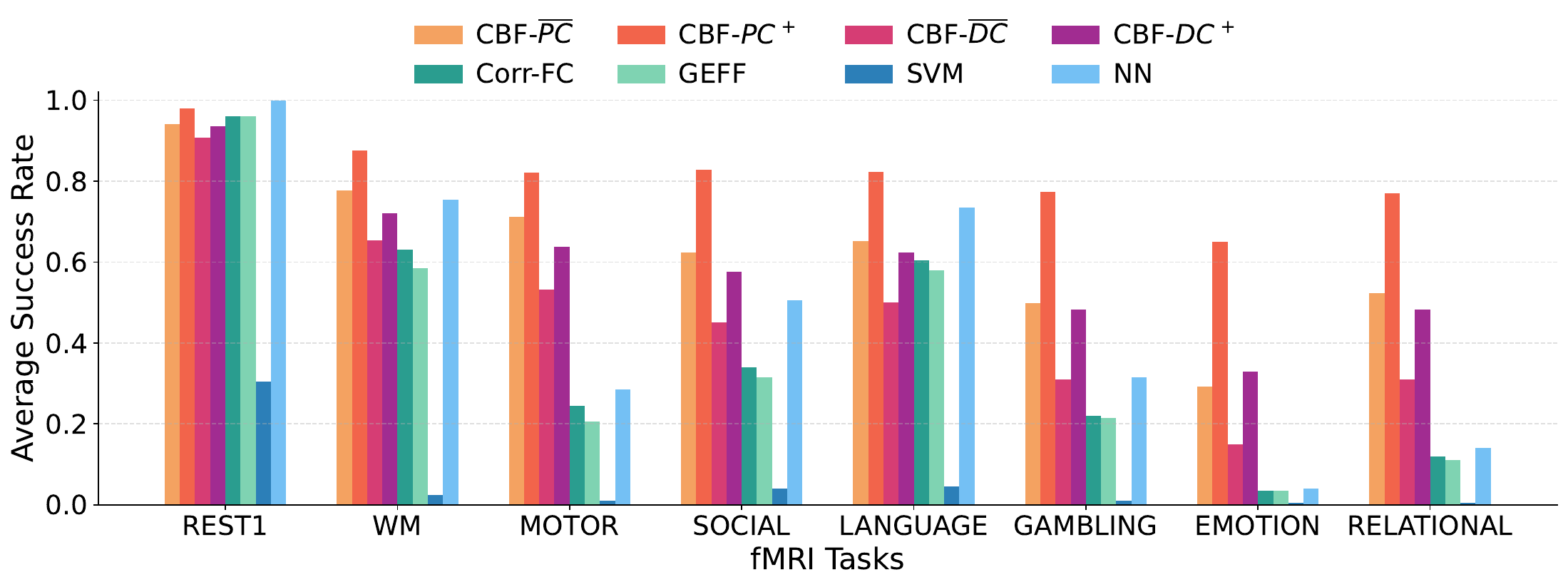}
    \caption{$S=200$}
    \label{fig:within_200}
\end{subfigure}

\vspace{2mm}

\begin{subfigure}[t]{0.49\textwidth}
    \centering
    \includegraphics[width=\linewidth]{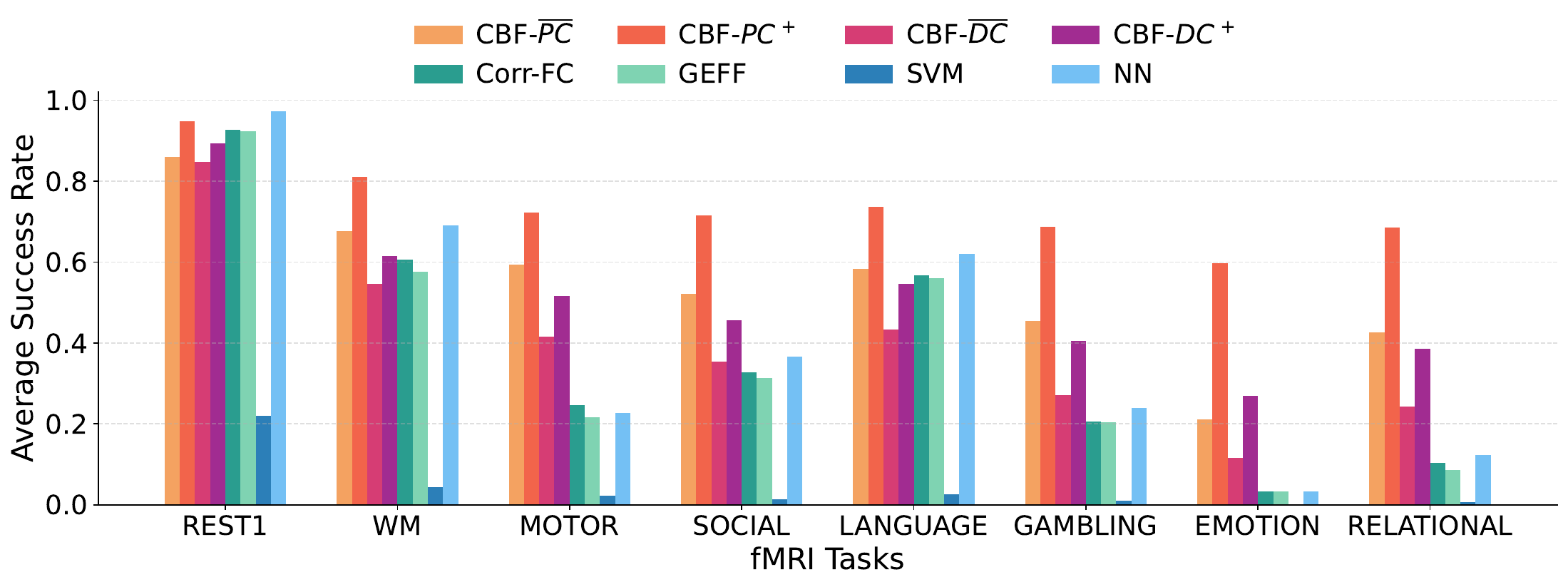}
    \caption{$S=300$}
    \label{fig:within_300}
\end{subfigure}
\hfill
\begin{subfigure}[t]{0.49\textwidth}
    \centering
    \includegraphics[width=\linewidth]{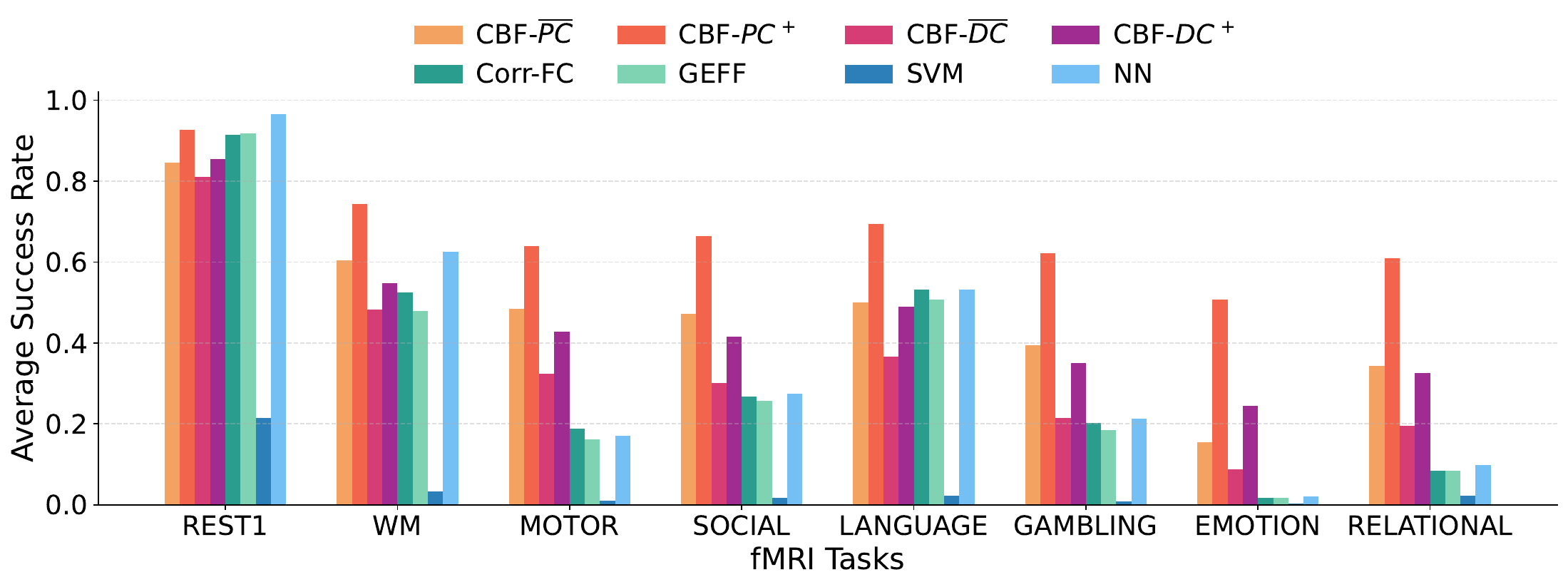}
    \caption{$S=400$}
    \label{fig:within_400}
\end{subfigure}

\caption{Within-task fingerprinting results across different methods for sample sizes (a) $S=200$, (b) $300$, and (c) $400$.}
\label{fig:within_large}
\end{figure*}

\begin{figure*}[!ht]
\centering
\includegraphics[width=\textwidth]{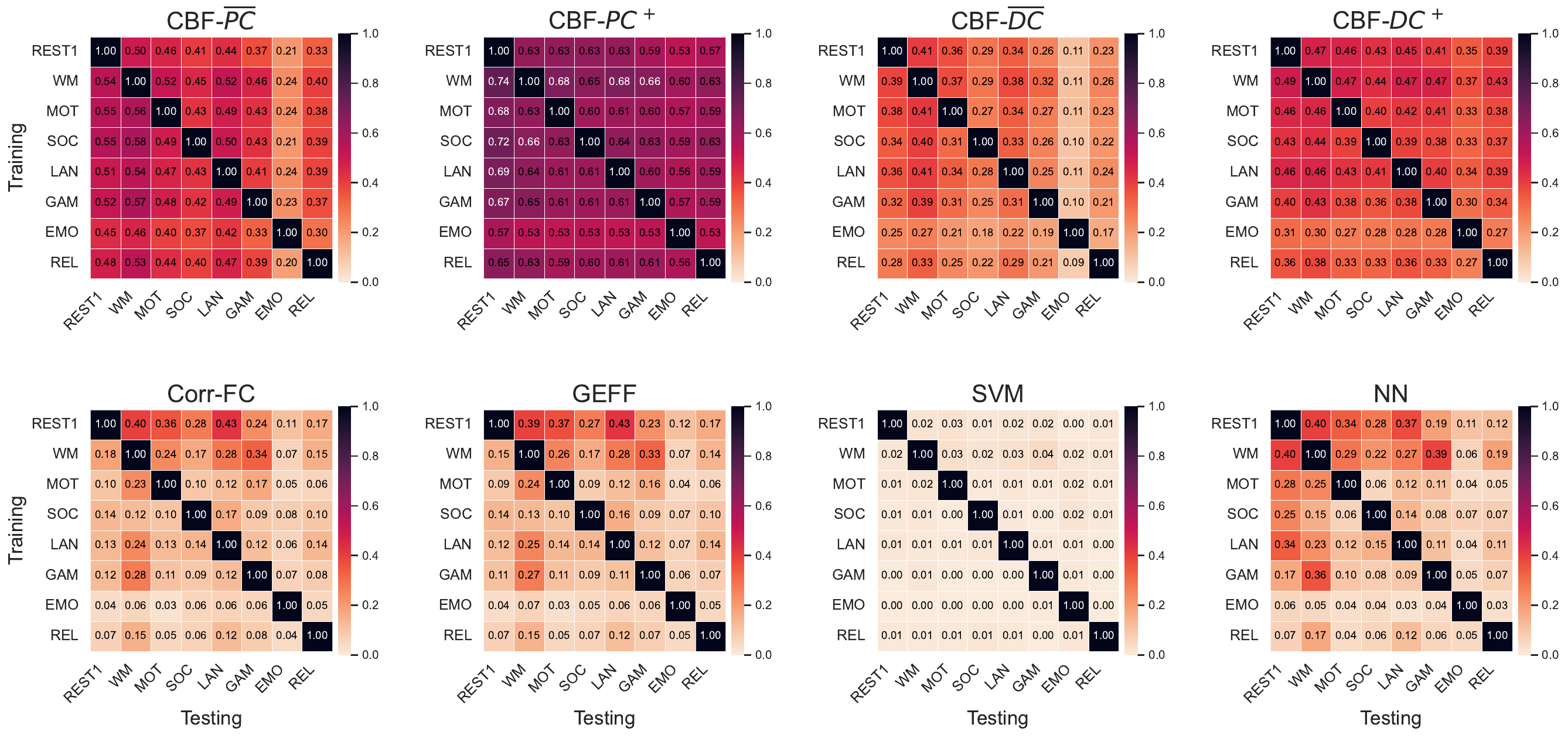}
\caption{Between-task fingerprinting results across different methods for a sample size of $S=400$. Results for $S=200$, and $300$ are provided in the Supplementary Materials.}
\label{fig:between_400}
\end{figure*}

 

\begin{figure*}[!ht]
\centering
\captionsetup[subfigure]{labelformat=empty}
\begin{subfigure}[t]{0.24\textwidth}
\centering
\includegraphics[width=\linewidth]{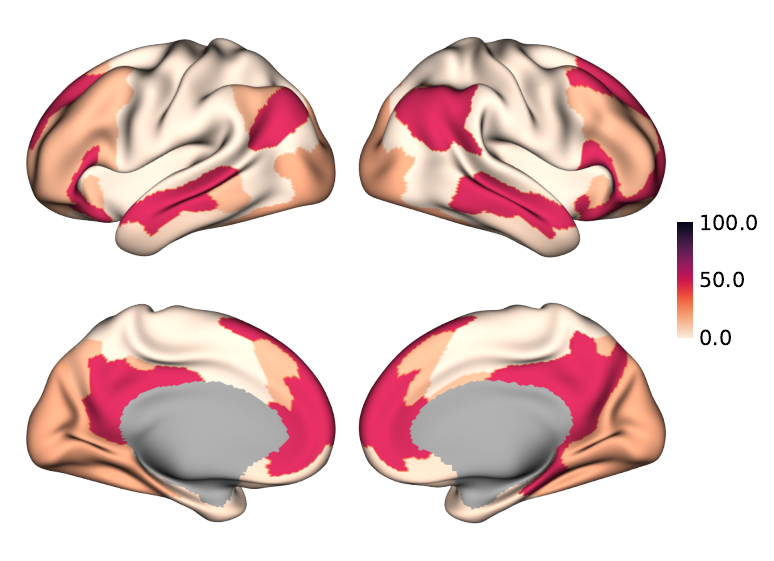}
\caption{Rest-DCM}
\label{Rest_DCM}
\end{subfigure}
\begin{subfigure}[t]{0.24\textwidth}
\centering
\includegraphics[width=\linewidth]{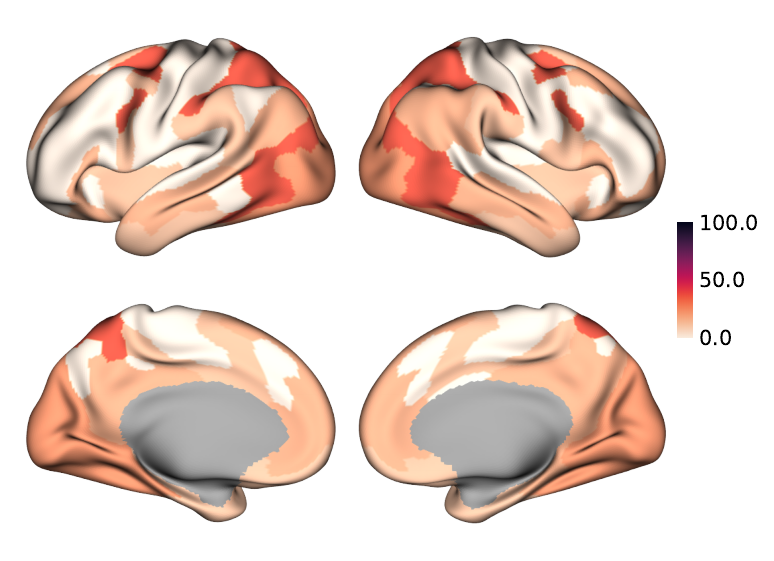}
\caption{WM-DCM}
\label{WM_DCM}
\end{subfigure}
\begin{subfigure}[t]{0.24\textwidth}
\centering
\includegraphics[width=\linewidth]{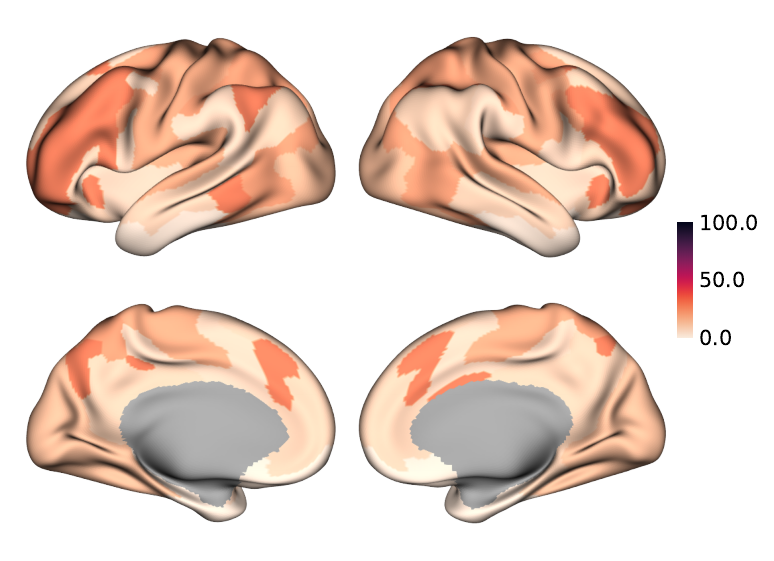}
\caption{MOT-DCM}
\label{MOT_DCM}
\end{subfigure}
\begin{subfigure}[t]{0.24\textwidth}
\centering
\includegraphics[width=\linewidth]{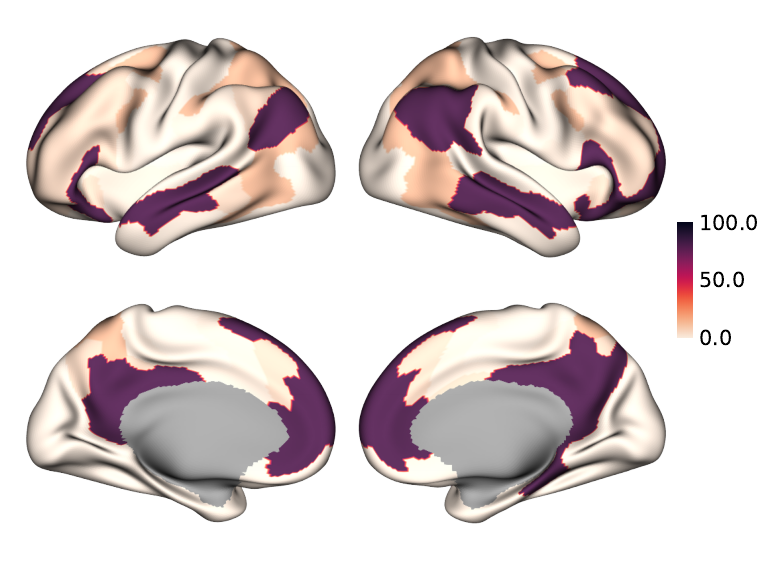}
\caption{SOC-DCM}
\label{SOC_DCM}
\end{subfigure}

\vspace{0.2cm}

\begin{subfigure}[t]{0.24\textwidth}
\centering
\includegraphics[width=\linewidth]{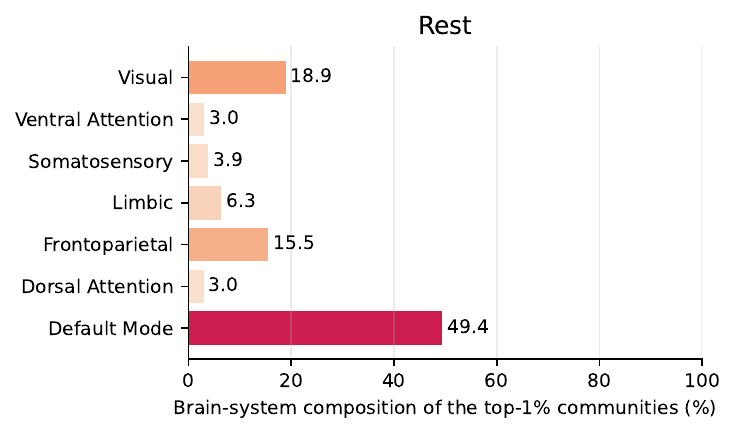}
\end{subfigure}
\begin{subfigure}[t]{0.24\textwidth}
\centering
\includegraphics[width=\linewidth]{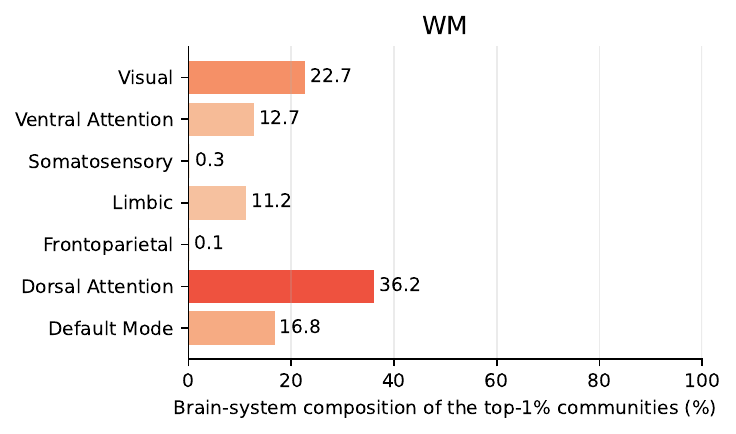}
\end{subfigure}
\begin{subfigure}[t]{0.24\textwidth}
\centering
\includegraphics[width=\linewidth]{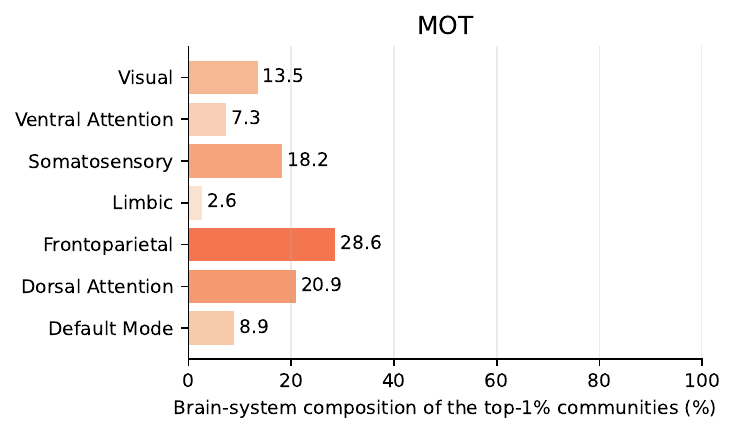}
\end{subfigure}
\begin{subfigure}[t]{0.24\textwidth}
\centering
\includegraphics[width=\linewidth]{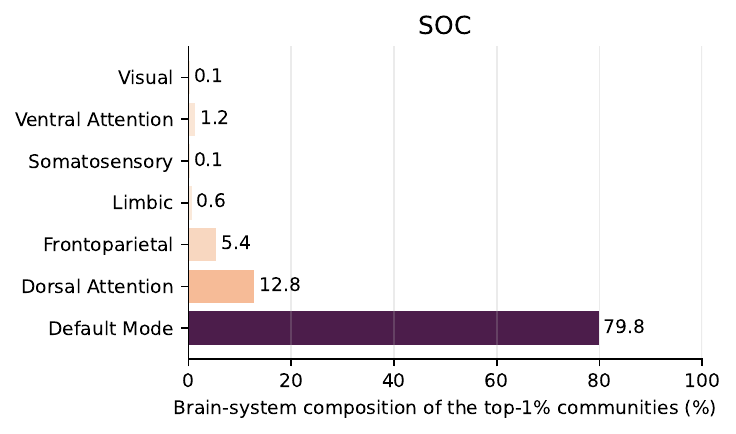}
\end{subfigure}

\vspace{0.5cm}

\begin{subfigure}[t]{0.24\textwidth}
\centering
\includegraphics[width=\linewidth]{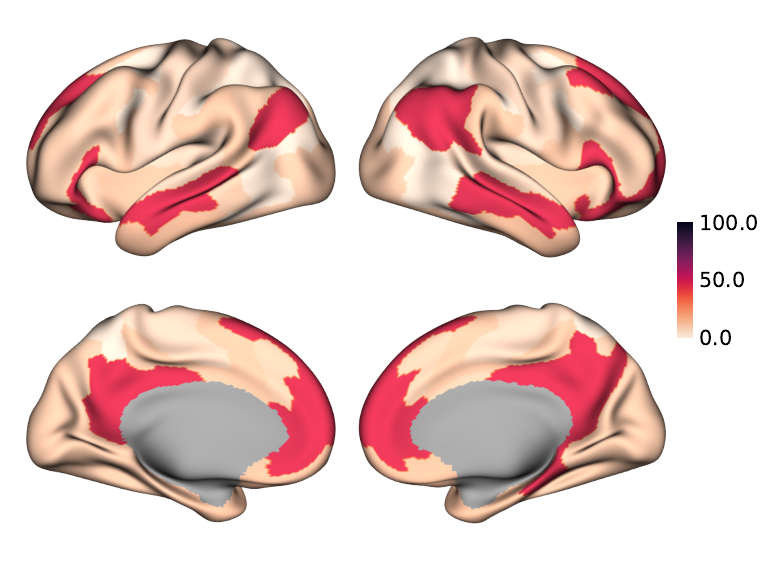}
\caption{LAN-DCM}
\label{LAN_DCM}
\end{subfigure}
\begin{subfigure}[t]{0.24\textwidth}
\centering
\includegraphics[width=\linewidth]{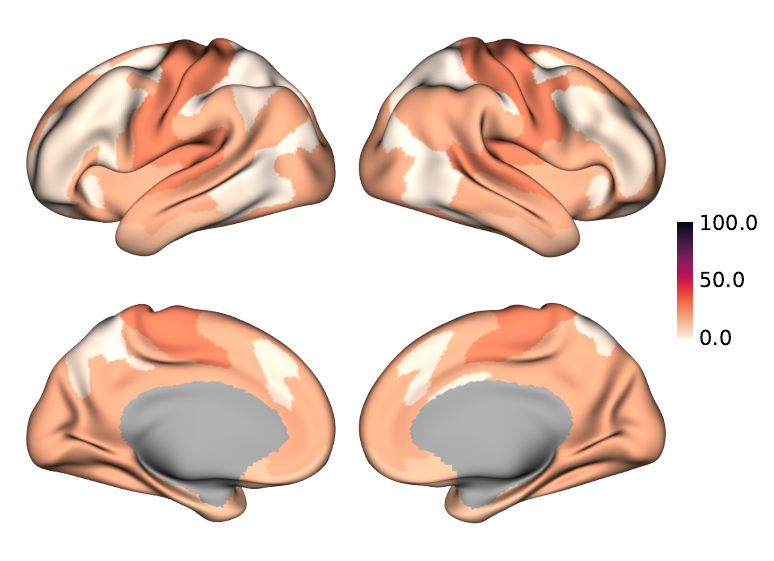}
\caption{GAM-DCM}
\label{GAM_DCM}
\end{subfigure}
\begin{subfigure}[t]{0.24\textwidth}
\centering
\includegraphics[width=\linewidth]{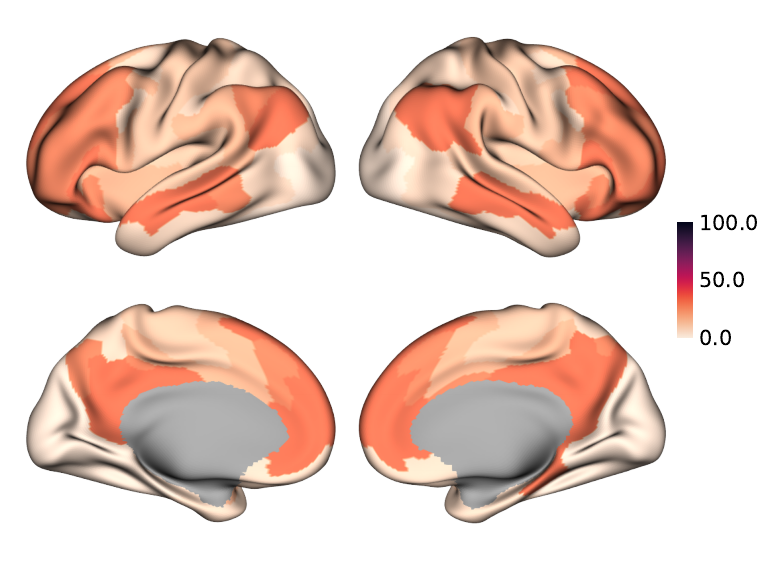}
\caption{EMO-DCM}
\label{EMO_DCM}
\end{subfigure}
\begin{subfigure}[t]{0.24\textwidth}
\centering
\includegraphics[width=\linewidth]{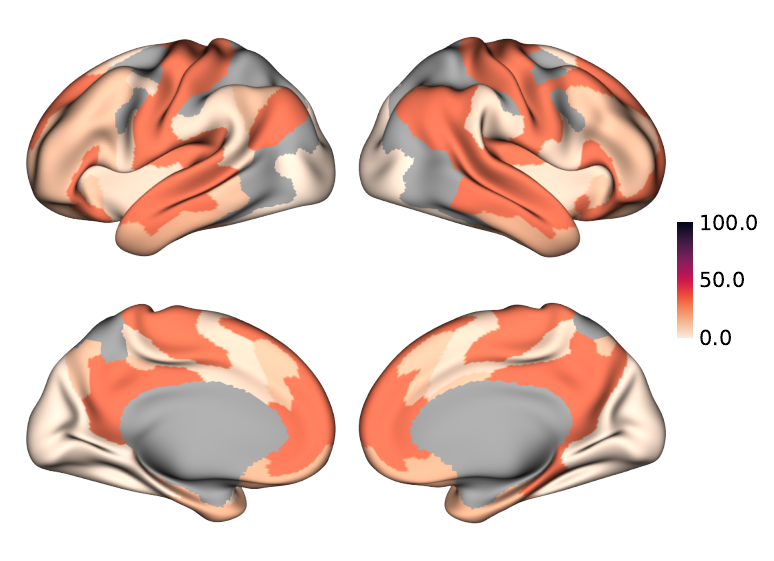}
\caption{REL-DCM}
\label{REL_DCM}
\end{subfigure}

\vspace{0.2cm}

\begin{subfigure}[t]{0.24\textwidth}
\centering
\includegraphics[width=\linewidth]{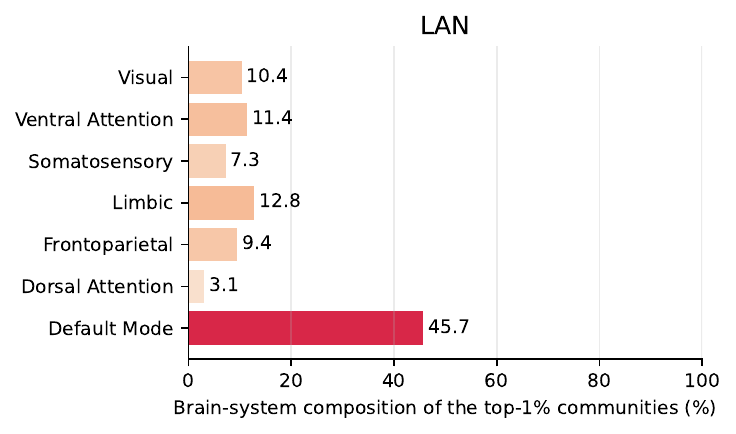}
\end{subfigure}
\begin{subfigure}[t]{0.24\textwidth}
\centering
\includegraphics[width=\linewidth]{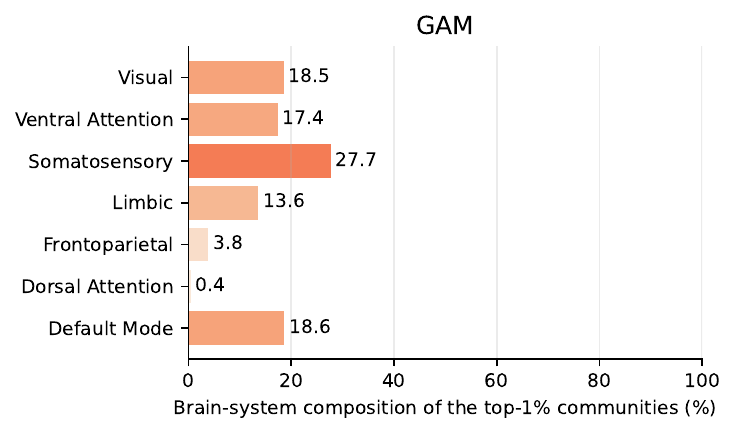}
\end{subfigure}
\begin{subfigure}[t]{0.24\textwidth}
\centering
\includegraphics[width=\linewidth]{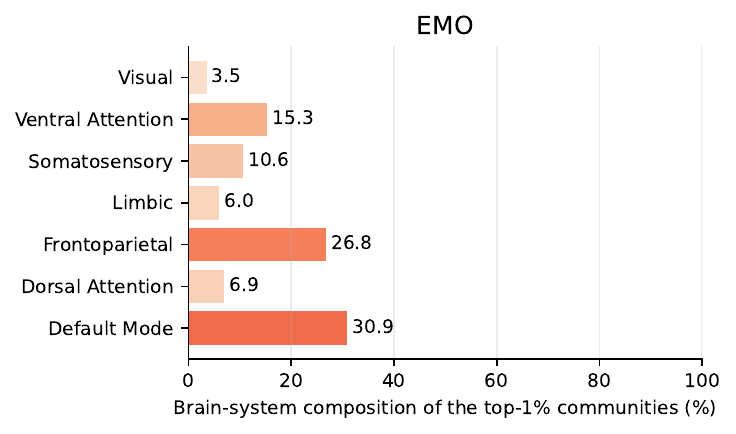}
\end{subfigure}
\begin{subfigure}[t]{0.24\textwidth}
\centering
\includegraphics[width=\linewidth]{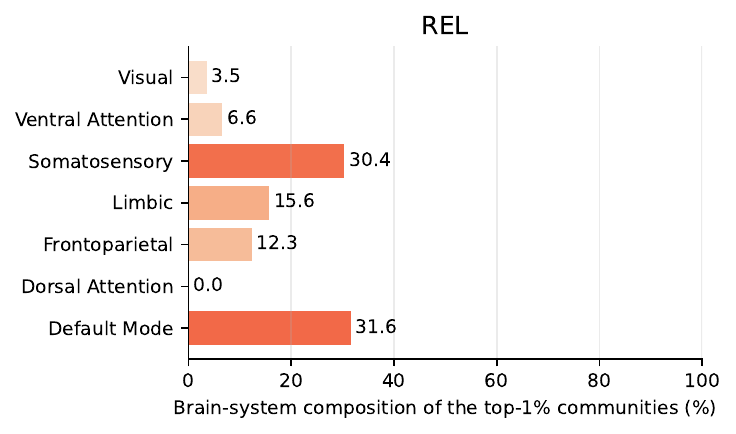}
\end{subfigure}

\caption{Discriminative connectivity maps illustrating the spatial distribution of the top-1\% discriminative community-based fingerprints (first and third rows), with the corresponding contributions of the seven Yeo functional brain systems (second and fourth rows) for each HCP task ($S=400$).}
\label{fig:Top1_400}
\end{figure*}

\section{Multiscale Community-Based Functional Connectome Fingerprinting}\label{sec:method}
The goal of this study is to extract community-based fingerprints (CBFs) that capture the unique mesoscale organization of each subject's functional connectome (Section ~\ref{CBF_ext}). To this end, the functional connectivity networks of a group of $S$ subjects are modeled as the layers of a signed multilayer network, $\{\mathcal{G}_{1}^{\pm},\ldots,\mathcal{G}_{S}^{\pm}\}$, with corresponding positive (negative) adjacency matrices $\matr{A}_{s}^{\pm}\in\mathbb{R}^{N\times N}$, $s\in\{1,\ldots,S\}$, whose entries represent the pairwise Pearson correlation coefficients between brain regions.
A signed multilayer modularity optimization framework is then applied to generate a set of $M$ multiscale partitions,
$\mathcal{P}=\{P_1,\ldots, P_M\}$,
where each partition $P_m$ consists of $K_m$ communities corresponding to the community assignments of all brain regions across all subjects $S$ (Section~\ref{CD_SMM}).
For each partition $P_m$, a low-dimensional community feature vector,
$\alpha_s^{m,f}\in\mathbb{R}^{K_m\times1}$,
is extracted for each subject $s$ and feature type $f$ defined as the community-based fingerprint (Section~\ref{CBF_ext}).
 \\ 
\noindent \textbf{Definition 1:} \label{def1}\emph{A community-based fingerprint is a feature vector, $\alpha_{s}^{m,f} \in\mathbb{R}^{K_{m}\times 1}$, derived from the subject's community assignment corresponding to partition, $P_{m}$, using graph-theoretic feature, $f$.}

Thus, for each subject $s$ and each feature type $f$, one obtains $M$ CBFs corresponding to each partition $P_{m}, m\in \{1,2,\ldots,M\}$.
The resulting $M$ CBF vectors are then evaluated to identify the top-$L$ partitions, that provide the most discriminative subject representations (Section~\ref{Partition Selection}). 
The selected top-$L$ partitions are subsequently used to construct (i) the training CBFs, $\{\alpha_{s}^{f,l}\}_{s=1}^{S}$, $l=1,\ldots,L$,
from the same FC networks used for SMM optimization (Section~\ref{CD_SMM}), and (ii) the testing CBFs, $\{\tilde{\alpha}_{s}^{f,l}\}_{s=1}^{S}$, $l=1,\ldots,L$, from the corresponding testing (unseen) FC networks of the same group of subjects.
Finally, each testing fingerprint is matched to its corresponding training fingerprint for partition $P_m$ and feature type $f$ using a distance-based nearest-neighbor classifier (Section~\ref{Sub_iden}). An overview of the proposed  framework is illustrated in Fig.~\ref{fig:overview}.
\subsection{Multiscale Community Detection using Signed Multilayer Modularity} \label{CD_SMM}
\subsubsection{Signed Multilayer Modularity}
Given $S$ subjects with FC matrices represented by the corresponding positive (negative) adjacency matrices, $\matr{A}_{s}^{\pm}$, $s\in \{1, \ldots, S\}$, we propose the signed multilayer modularity as:
\begin{equation}
\begin{gathered} \arg\max_{{\matr{C}}}
Q^{\pm}(\matr{C}|\gamma,\omega;\matr{A}_s^{\pm},\matr{P}_s^{\pm}):= \sum_{s,r=1}^{S}\sum_{i,j=1}^{N}
\Big[(A_{ijs}^{\pm}- \gamma P_{ijs}^{\pm})\delta_{sr} +\omega_{jsr}\delta_{ij}\Big]\delta(C_{is},C_{jr}).
\label{eq:SMM}
\end{gathered}\end{equation}
\noindent 
This definition of SMM overcomes the limitations of traditional multilayer modularity \cite{Mucha2010} optimization by incorporating both correlated and anti-correlated activity, providing a robust tool for studying the variability of network organization across individuals. Since this optimization problem is NP hard, it does not have a closed form mathematical solution.
Instead, SMM is optimized using greedy algorithms such as the Louvain method \cite{blondel2008fast}, yielding community assignments across subjects simultaneously while accommodating inter-subject variability.

Rather than assuming the existence of a single optimal parameter pair, the proposed framework is based on the observation that different combinations of the modularity parameters $\{\gamma,\omega\}$ can reveal meaningful joint community structures at distinct topological scales. Accordingly, the Louvain algorithm is executed over a restricted rectangular region of the $\{\gamma,\omega\}$ parameter space, resulting in a set of $M$ candidate partitions $\{P_{1}, \ldots, P_{M}\}$, where each partition $P_{m}$ consists of $K_m$ disjoint communities $\{c_{1},\ldots, c_{K_{m}}\}$. 
We exclude extreme or uninformative partitions such as singleton communities, whole-network partitions, those identical across all layers, and those maximally dissimilar across layers, following the framework of \cite{betzel2017multi}.
\subsection{Fingerprint Extraction}
\label{CBF_ext}

Optimizing the SMM objective in \eqref{eq:SMM} yields $M$ candidate partitions, $\{P_m\}_{m=1}^{M}$. For each partition $P_m$ with $K_m$ communities $\{c_{1}, \ldots, c_{K_m}\}$, we construct a community-based fingerprint vector for each subject $s$ and feature type $f$, $\alpha_{s}^{f,m}\in\mathbb{R}^{K_m\times1}$. The ($k$)-th entry, $\alpha_{s}^{f,m}(k)$, corresponds to the value of the graph-theoretic feature $f$ computed for community $c_k$, thereby characterizing its topological role within the overall network architecture. In this work, we consider two main graph-theoretic features: (i) the participation coefficient (PC) and (ii) the diversity coefficient (DC)
\cite{garcia2018applications,guimera2005cartography,rubinov2011weight}.  
Each feature type is evaluated independently, resulting in $M$ sets of CBFs, one for each candidate partition and feature type.

\subsubsection{Participation coefficient}
The participation coefficient is an index of global inter-modular integration \cite{garcia2018applications}. 
For signed graphs, we compute two participation coefficient measures: the positive participation coefficient, $PC^+$, using only positive edges, and the total participation coefficient, $\overline{PC}$, using the entire signed network, thereby capturing the distinct roles of positive and negative edges.
The positive participation coefficient is defined as:

\begin{equation}
{PC}^{+}_{is}= 1-\sum_{k=1}^{K_m} p^{+}_{is}(c_{k}), \quad \text{s.t.} \quad  p^{+}_{is}(c_k)=\left(\frac{\mathcal{K}^{+}_{is}(c_{k})}{\mathcal{K}^{+}_{is}}\right),
\end{equation}
where $p^{+}_{is}(c_{k})$ is the ratio of the strength of positive edges from node $i$ to nodes in community $c_{k}$, to the total positive strength of node $i$ for subject $s$. Similarly, the participation coefficient with respect to negative edges, ${PC}_{is}^{-}$, can be computed.
The total  participation coefficient for positive and negative edges, $\overline{PC}_{is}$, is then computed as:
 \begin{equation}
\overline{PC}_{is}=PC^{+}_{is}-\frac{\mathcal{K}^{-}_{is}}{\mathcal{K}^{-}_{is}+\mathcal{K}^{+}_{is}}PC^{-}_{is},
 \end{equation}
emphasizing the role of positive edges more than the negative edges in community formation \cite{rubinov2011weight}.
Given $\{PC^{+}_{is}\}_{i=1}^{N}$ and $\{\overline{PC}_{is}\}_{i=1}^{N}$, one can quantify the average positive and total participation coefficients of each community, $c_k$, by averaging the participation coefficients within each community as: 
\begin{equation}
    \alpha_{s}^{1}(k)=\frac{1}{|c_{k}|}\sum_{i\in c_{k}}\overline{PC}_{is}.
\label{eq:bar_PC}
\end{equation}
\begin{equation}
    \alpha_{s}^{2}(k)=\frac{1}{|c_{k}|}\sum_{i\in c_{k}}PC^{+}_{is}.
    \label{eq:PC_p}
\end{equation}
\subsubsection{Diversity coefficient}
The diversity coefficient quantifies the regional connection diversity using Shannon's normalized entropy. 
We compute two diversity coefficient measures: the positive diversity coefficient, $DC^+$, using only positive edges, and the total diversity coefficient, $\overline{DC}$, using the entire signed network.
The positive diversity coefficient  can be computed as:

\begin{equation}
DC^{+}_{is}=-\frac{1}{\log(K_m)}\sum_{k=1}^{K_m} p^{+}_{is}(c_k)\log (p^{+}_{is}(c_k)).
\end{equation}

Similarly, the diversity coefficient with respect to negative edges, $DC_{is}^{-}$, and the total diversity coefficient, $\overline{DC}_{is}$, are computed. 
Given $\{DC^{+}_{is}\}_{i=1}^{N}$ and $\{\overline{DC}_{is}\}_{i=1}^{N}$, the positive and total diversity of each community, $c_k$, can be quantified as:
\begin{equation}
    \alpha^{3}_{s}(k)=\frac{1}{|c_{k}|}\sum_{i\in c_{k}}\overline{DC}_{is}.
    \label{eq:bar_DC}
\end{equation}
\begin{equation}
    \alpha^{4}_{s}(k)=\frac{1}{|c_{k}|}\sum_{i\in c_{k}}DC_{is}^{+}.
    \label{eq:DC_p}
\end{equation}


\vspace{-0.3in}
\subsection{Partition Selection}
\label{Partition Selection}
For each feature type, $f$, the resulting $M$ CBFs extracted in Section \ref{CBF_ext}, are evaluated to identify the partitions that provide the most discriminative subject representations. To this end, we compute the mean pairwise Euclidean distance (MPED) across all subjects for each partition:
\begin{equation}\phi^{f,m}= \frac{2}{S(S-1)} \sum_{s=1}^{S-1} \sum_{s'=s+1}^{S}  \lVert \alpha_{s}^{f,m} - \alpha_{s'}^{f,m} \rVert_{2}.
\label{eq:phi}
\end{equation}
\noindent 
The partitions are ranked in descending order according to their MPED score, $\phi^{f,m}$. Rather than relying on a single highest-ranked partition, which may fail to capture the full spectrum of mesoscale network organization, the proposed framework selects the top-$L$ partitions. 
This approach retains multiple highly discriminative partitions representing complementary organizational patterns across different topological scales. The selected partitions are evaluated independently for fingerprinting while also providing multiple network representations for the interpretability analysis described in Section~\ref{C-ICC}.

The number of top-ranked partitions was selected by evaluating MPED as a function of the number of rank-ordered partitions. The smallest number of top-ranked partitions, denoted by $L$, that attain at least $99\%$ of the maximum MPED are selected (see Supplementary Section S.I.).



\vspace{-0.1in}
\subsection{Subject Identification}
\label{Sub_iden}
The resulting top-$L$ partitions (Section ~\ref{Partition Selection}) are then used to construct the training CBFs, $\{\alpha_{s}^{f,l}\}_{s=1}^{S}$, from the training FC networks used for SMM optimization in Section~\ref{CD_SMM}. The same partitions are subsequently applied to the corresponding testing (unseen) FC networks of the same subjects to construct the testing CBFs, $\{\tilde{\alpha}_{s}^{f,l}\}_{s=1}^{S}$. This ensures that both the training and testing scans are represented in a common feature space, enabling direct comparison for subject identification.

For each selected partition $P_{l}$ and feature type $f$, the testing CBFs, $\{\tilde{\alpha}_{s}^{f,l}\}_{s=1}^{S}$, is matched against the corresponding training CBFs, $\{\alpha_{s}^{f,l}\}_{s=1}^{S}$, and subjects are identified using a distance-based nearest-neighbor classifier. Specifically, the predicted subject index, $s^{*}$, is determined as the index of the training CBF vector, $\alpha_{s'}^{f,l}$, that minimizes the Euclidean distance to the testing CBF vector, $\tilde{\alpha}_{s}^{f,l}$:
\begin{equation}
\label{eq:min_dis}
s^{*} = \arg\min_{s'} \, \lVert \tilde{\alpha}^{f,l}_{s} - \alpha^{f,l}_{s'} \rVert_{2}.
\end{equation}
\noindent 
The subject identification success rate is defined as the percentage of correctly identified subjects relative to the total number of subjects.

\vspace{-0.15in}
\subsection{Community-based fingerprint interpretability}
\label{C-ICC}
Finally, we evaluate the reliability of the detected communities in the top-ranked partitions by computing the intra-class correlation coefficient (ICC) between the corresponding training and testing CBF pairs \cite{koo2016guideline,amico2018quest,sareen2021exploring}. 
For a given group of subjects, we define a \emph{community-based ICC matrix (C-ICC)}, $\mathcal{J}\in\mathbb{R}^{K_m\times K_m}$, whose $(k,l)$-th entry quantifies the test--retest reliability across subjects between the community-level feature of community $c_k$ in the training CBF and that of community $c_l$ in the testing CBF.
We define the \emph{self-reliability} ($\mathcal{S}$) of a community as the ICC value of the community, $c_k$, with itself such that $\mathcal{S}_{c_{k}}= \mathcal{J}_{kk}$. This quantifies how stable a community is across subjects and sessions. Higher self-reliability values indicate greater stability of the extracted community-based fingerprints. 
In contrast, the ICC values between a community and all other communities quantify its \emph{relational similarity} ($\mathcal{R}$), capturing undesired overlap or redundancy with other communities. This is computed as the average of the off-diagonal entries of $\mathcal{J}$ matrix for community, $c_{k}$:
\begin{equation}
        \mathcal{R}_{c_{k}} = \frac{1}{\,K_m-1\,} \sum_{l \neq k}^{K_m} \mathcal{J}_{kl}.
\end{equation}
In order to identify communities that simultaneously have high self-reliability, $\mathcal{S}_{c_{k}}$, and low relational similarity, $\mathcal{R}_{c_{k}}$, we define a 
distinctiveness score ($\mathcal{D}$) for each community:
\begin{equation}
        \mathcal{D}_{c_{k}} = \mathcal{S}_{c_{k}} - \mathcal{R}_{c_{k}}.
\end{equation}
Based on the distinctiveness score, communities are ranked in descending order, where communities with larger $\mathcal{D}_{c_{k}}$ values are more reliable and more distinct from others.

To provide an interpretation of the highly distinctive communities, the highest-ranked communities based on $\mathcal{D}_{c_k}$ are selected, and their constituent ROIs are identified by analyzing their spatial composition within the corresponding joint community assignment matrix, $\matr{C}_l \in \{1, \dots, K\}^{N \times S}$.
The joint community assignment enables characterization along two complementary dimensions. Along the subject dimension $S$, the identified ROIs can be examined for an individual subject, a subgroup of subjects, or an entire cohort. Along the spatial dimension $N$, they can be analyzed at the nodal level or grouped into functional brain systems to quantify system-level contributions. 
In this work, we focus on the cohort-level, system-level representation, in which ROIs are mapped to the seven large-scale functional brain systems defined by the Yeo atlas \cite{yeo2011organization} to quantify the spatial distribution of the most discriminative communities across subjects. 
Specifically, a $N\times S$ binary community membership matrix is constructed, where entries, i.e., brain regions and subjects, associated with the selected communities are assigned a value of one and all remaining entries are assigned a value of zero. The contribution of each brain system is then quantified as the percentage of non-zero entries associated with ROIs (nodes) belonging to that system across subjects.
The resulting spatial representations, referred to as \emph{Discriminative Connectivity Maps}, highlight the functional brain systems that contribute most strongly to individual identification.


\subsection{Computational Complexity}
The computational complexity of the proposed framework is dominated by the optimization of the signed multilayer modularity. For a signed multilayer network with $S$ subjects and $N$ nodes, the multilayer Louvain optimization has an expected cost of $\mathcal{O}(S N^{2})$ per partition. Given $M$ partitions generated across different $(\gamma,\omega)$ parameter settings, the overall computational complexity of the framework is therefore $\mathcal{O}(M S N^{2})$.

\section{Experimental Results}
In this section, we evaluate the proposed framework in two complementary settings, i.e., within-task and between-task fingerprinting, to evaluate the reproducibility of the results across various fMRI conditions and different subject groups \cite{gao2021smooth}.

\subsection{fMRI Dataset and Preprocessing}
The proposed framework was evaluated using multilayer FC networks derived from minimally preprocessed, FIX-cleaned 3-Tesla resting-state fMRI (rsfMRI) and seven task fMRI (tfMRI) datasets, including emotion processing (EMO), gambling (GAM), language (LAN), motor (MOT), relational processing (REL), social cognition (SOC), and working memory (WM) tasks, obtained 
 as part of the Human Connectome Project (HCP) S1200 release \cite{van2013wu} \footnote{\href{https://db.humanconnectome.org/}{The Human Connectome Project (HCP).}}. Only subjects who completed both right–left (RL) and left–right (LR) phase-encoding scans for all eight fMRI conditions (REST1 and seven tasks) were included, yielding a final sample of 810 subjects.
We apply further fMRI preprocessing steps, including parcellation, linear detrending, band-pass filtering (0.008–0.08 Hz), confound regression, and z-score standardization of the extracted BOLD signals. FCs are estimated using Pearson’s correlation between time series derived from the Multi-Modal Parcellation (MMP) with 360 brain regions (nodes) \cite{glasser2016multi} and matched to Yeo’s 7-network parcellation \cite{yeo2011organization}.

\subsection{Baselines}
We compare the performance of the proposed framework to state-of-the-art fingerprinting algorithms, including correlation-based functional connectivity (Corr-FC) \cite{finn2015functional}, graph embedding for functional fingerprinting (GEFF) \cite{abbas2020geff}, a support-vector machine (SVM) classifier, and a multi-layer perceptron neural network (NN) classifier \cite{hannum2023high}.
Throughout this section, CBF-$PC$, CBF-$PC^{+}$, CBF-$DC$, and CBF-$DC^{+}$ willl refer to the proposed community-based fingerprint corresponding to $PC$, $PC^{+}$, $DC$, and $DC^{+}$ features, respectively.

\subsection{Within-task fingerprinting}
\label{within_task}

Using the RL scans for a given group of subjects, (a) we generate a large number of candidate partitions ($M=1,000$) using SMM over a range of $\gamma \in [0,2]$ and $\omega \in [0,2]$ (Section~\ref{CD_SMM}); 
(b) for each partition, CBF vectors are extracted for all subjects from the RL scan of each task using the four feature types defined in Eqs.~\eqref{eq:bar_PC}--\eqref{eq:DC_p} (Section~\ref{CBF_ext});
(c) we compute the MPED score for each partition and rank the partitions  as described in Section~\ref{Partition Selection}, to determine a set of $L$ discriminative partitions  for fingerprinting; (d) 
The selected top-$L$ partitions are used to construct the training and testing CBFs from the RL and corresponding LR scans, respectively.
Subject identity is then determined using the distance-based classifier defined in Section~\ref{Sub_iden}.

\subsubsection{Fingerprinting analysis on a small-scale dataset}
We first evaluate the proposed pipeline on a small sample size of $S=100$. For this analysis, we randomly select ten independent subject groups, each with a sample size of $S=100$, and the entire pipeline is repeated for each feature type and task. Evaluating multiple randomly selected subject groups allows us to assess the robustness of the proposed framework to subject selection and reduces the possibility that the observed results are driven by a particular group of individuals.

The average identification accuracy across the ten repetitions is reported in Fig.~\ref{fig:within_100}. From this figure, it can be seen that the proposed framework achieves consistently high identification performance across all eight fMRI tasks. Among the four proposed features, CBF-$PC^+$ consistently provides the highest accuracy, followed by CBF-$PC$, whereas CBF-$DC^+$ and CBF-$DC$ yield lower but competitive performance. Compared with the baseline methods, the proposed approach achieves the highest identification accuracy for most tasks, while NN slightly outperforms CBF-$PC^+$ for the REST1 and LAN tasks.

\emph{Statistical Analysis:}
To assess the robustness of fingerprinting performance to subject selection and determine whether identification accuracy varies across fingerprinting methods and fMRI tasks, we performed a two-way repeated-measures ANOVA with method and fMRI task as fixed within-subject factors, treating the ten independently selected random subject groups as repeated observations. Greenhouse-Geisser (GG) corrected $p$-values were reported to account for potential violations of the sphericity assumption \cite{geisser1958extension,huynh1976estimation}. Significant main effects and interactions were further examined using Bonferroni-corrected post-hoc pairwise comparisons.

As summarized in Table~\ref{tab:anova_2way}, significant main effects of fMRI task ($F(7,63)=1181.6$, $p$-value (GG)=$1.08\times10^{-31}$) and method ($F(7,63)=5834.7$, $p$-value (GG)$=6.48\times10^{-28}$), as well as a significant fMRI task $\times$ method interaction ($F(49,441)=180.6$, $p$-value (GG)$=1.74\times10^{-30}$), indicate that fingerprinting performance depends on both the experimental condition and the fingerprinting method. Bonferroni-corrected post-hoc comparisons (Table~\ref{tab:posthoc_anova_2way}) further showed that the proposed framework based on the $PC^{+}$ metric significantly outperformed all competing methods, including $PC$, $DC^{+}$, $DC$, Corr-FC, GEFF, NN, and SVM (all adjusted $p<0.001$). Among the proposed CBF features, the performance ranking was \mbox{CBF-$PC^{+}$} $>$ \mbox{CBF-$PC$} $>$ \mbox{CBF-$DC^{+}$} $>$ \mbox{CBF-$DC$}, with all pairwise differences reaching statistical significance.

\emph{Sensitivity Analysis of Top-$L$ Partitions:} 
We evaluate MPED as a function of the number of top-ranked partitions and select the smallest $L$ achieving at least $99\%$ of the maximum MPED. This analysis resulted in $L=10$, which was used throughout all experiments (Fig.~S1 and Fig.~S2). Detailed sensitivity analysis results are provided in the Supplementary Material.

\emph{Sensitivity Analysis of the SMM Parameters:}
To further assess the robustness of the proposed framework, we examine the effects of the SMM parameters $\gamma$ and $\omega$ on both partition discriminability, quantified by MPED, and fingerprinting accuracy across all tasks. The analysis shows that the proposed framework does not depend on a unique optimal parameter pair and maintains high discriminability and identification performance across a range of $(\gamma,\omega)$ values (Fig.~S3). Detailed sensitivity analysis results are provided in the Supplementary Material.

\subsubsection{Fingerprinting analysis on a large-scale dataset}
Next, we evaluate the proposed framework on large-scale datasets using sample sizes of $S=200, 300$, and $400$. For this evaluation, we focus on only one randomly selected subject group for each sample size as the computational complexity of SMM increases substantially with $S$.
This large-scale analysis aims to assess the scalability of the proposed framework, evaluate the effect of increasing sample size on identification performance, and determine whether the extracted community-based fingerprints exhibit consistent and generalizable patterns as the cohort size increases.

As shown in Fig.~\ref{fig:within_large}, subject identification accuracy decreases with increasing $S$. The proposed framework exhibits robustness to increasing sample size, maintaining relatively high identification accuracy across both sample sizes and fMRI tasks. Among the proposed features, CBF-$PC^{+}$ performs best across  both increasing sample sizes and diverse fMRI conditions. In contrast, the baseline methods show greater sensitivity to sample size and task condition. SVM exhibits the lowest overall performance and limited robustness to both increasing sample size and changes in task condition. Although the NN method is comparable to CBF-$PC^{+}$ for the resting-state task across sample sizes, its performance drops drastically for task-based fMRI, whereas CBF-$PC^{+}$ remains robust against growing sample sizes across tasks.
Corr-FC and GEFF also perform relatively well for resting-state fMRI but show considerable performance degradation under task-based conditions.

\emph{Statistical Analysis:} To evaluate whether identification accuracy varies across fingerprinting methods, sample sizes, and fMRI tasks, we performed a three-way ANOVA with method, sample size, and fMRI task as fixed factors. Since only one randomly selected subject group was evaluated for each sample size, the analysis was performed using a standard factorial ANOVA. Significant effects were further examined using Bonferroni-corrected post-hoc pairwise comparisons.

As summarized in Table~\ref{tab:anova_3way}, the three-way ANOVA revealed significant main effects of fMRI task ($F(7,98)=2531.54$, $p=1.03\times10^{-107}$), sample size ($F(2,98)=373.31$, $p=1.46\times10^{-46}$), and method ($F(7,98)=2269.96$, $p=2.09\times10^{-105}$). Significant interactions were also observed between method and sample size ($F(14,98)=13.77$, $p=1.90\times10^{-17}$), method and fMRI task ($F(49,98)=77.45$, $p=1.68\times10^{-60}$), and sample size and fMRI task ($F(14,98)=4.13$, $p=1.27\times10^{-5}$). These results indicate that identification performance depends on the fingerprinting approach, the cohort size, and the fMRI task.
Bonferroni-corrected post-hoc comparisons (Table~\ref{tab:posthoc_anova_3way}) showed that the proposed framework based on the $PC^{+}$ metric significantly outperformed all competing methods, including CBF-PC, CBF-DC$^{+}$, CBF-DC, Corr-FC, GEFF, NN, and SVM (all adjusted $p<0.05$). 
\begin{table}[t]
\centering
\caption{
Results of the two-way repeated-measures ANOVA for sample size $S$=100.}
\setlength{\tabcolsep}{2pt}
\label{tab:anova_2way}
\begin{tabular}{lcccc}
\hline
\textbf{Effect} & \textbf{DF} & \textbf{F-value} & \textbf{$p$-value} & \textbf{$p$-value (GG)} \\
\hline
fMRI task & $7,63$ & 1181.60 & $2.79\times10^{-64^{*}}$ & $1.08\times10^{-31^{*}}$ \\
Method & $7,63$ & 5834.68 & $4.89\times10^{-86^{*}}$ & $6.48\times10^{-28^{*}}$ \\
fMRI task $\times$ Method & $49,441$ & 180.61 & $1.47\times10^{-260^{*}}$ & $1.74\times10^{-30^{*}}$ \\
\hline
\end{tabular}
\par\vspace{2pt}
\noindent
\parbox{\linewidth}{\centering\footnotesize 
\textit{Abbreviations:} DF = degrees of freedom; GG = Greenhouse--Geisser correction. \\ Asterisks ($^{*}$) indicate statistical significance at $p<0.05$.}
\end{table}

\begin{table}[t]
\centering
\setlength{\tabcolsep}{2pt}
\caption{Bonferroni-corrected post-hoc comparisons for the method factor, for sample size $S$=100.}
\label{tab:posthoc_anova_2way}
\begin{subtable}{\columnwidth}
\centering
\caption{Comparison among the proposed CBF features.}
\begin{tabular}{lccc}
\hline
\textbf{Comparison} & \textbf{Mean Difference} & \textbf{Std. Err.} & \textbf{Adj. $p$-value} \\
\hline
CBF-$PC^+$ vs. CBF-$PC$   & 0.1492 & 0.0031 & $7.21\times10^{-8^{*}}$ \\
CBF-$PC^+$ vs. CBF-$DC^+$ & 0.1700 & 0.0033 & $7.21\times10^{-8^{*}}$ \\
CBF-$PC^+$ vs. CBF-$DC$   & 0.3103 & 0.0048 & $7.21\times10^{-8^{*}}$ \\
CBF-$PC$ vs. CBF-$DC^+$   & 0.0208 & 0.0014 & $1.63\times10^{-6^{*}}$ \\
CBF-$PC$ vs. CBF-$DC$     & 0.1611 & 0.0023 & $7.21\times10^{-8^{*}}$ \\
CBF-$DC^+$ vs. CBF-$DC$   & 0.1404 & 0.0018 & $7.21\times10^{-8^{*}}$ \\
\hline
\end{tabular}
\end{subtable}
\vspace{3mm}
\begin{subtable}{\columnwidth}
\centering
\caption{Comparison of the best-performing CBF feature (CBF-$PC^+$) with baseline methods.}
\begin{tabular}{lccc}
\hline
\textbf{Comparison} & \textbf{Mean Difference} & \textbf{Std. Err.} & \textbf{Adj. $p$-value} \\
\hline
CBF-$PC^+$ vs. Corr-FC & 0.4520 & 0.0035 & $7.21\times10^{-8^{*}}$ \\
CBF-$PC^+$ vs. GEFF    & 0.4725 & 0.0034 & $7.21\times10^{-8^{*}}$ \\
CBF-$PC^+$ vs. NN      & 0.2680 & 0.0038 & $7.21\times10^{-8^{*}}$ \\
CBF-$PC^+$ vs. SVM     & 0.8188 & 0.0031 & $7.21\times10^{-8^{*}}$ \\
\hline
\end{tabular}
\par\vspace{2pt}
\noindent
\parbox{\linewidth}{\centering \footnotesize \textit{Abbreviations}: Std. Err., standard error; Adj. $p$-value, Bonferroni-adjusted $p$-value.}
\end{subtable}
\end{table}

\begin{table}[t]
\centering
\setlength{\tabcolsep}{2pt}
\caption{Results of the
three-way ANOVA for sample size $S$=200, 300, and 400.}
\label{tab:anova_3way}
\begin{tabular}{lccc}
\hline
\textbf{Effect} & \textbf{DF} & \textbf{F-value} & \textbf{$p$-value} \\
\hline
fMRI task                          & $7,98$  & 2531.54 & $1.03\times10^{-107}$ \\
Sample Size                        & $2,98$  & 373.31  & $1.46\times10^{-46}$ \\
Method   & $7,98$  & 2269.96 & $2.09\times10^{-105}$ \\
Method $\times$ Sample Size        & $14,98$ & 13.77   & $1.90\times10^{-17}$ \\
Method $\times$ fMRI task          & $49,98$ & 77.45   & $1.68\times10^{-60}$ \\
Sample Size $\times$ fMRI task     & $14,98$ & 4.13    & $1.27\times10^{-5}$ \\
\hline
\end{tabular}
\end{table}

\begin{table}[t]
\centering
\setlength{\tabcolsep}{2pt}
\caption{Bonferroni-corrected post-hoc comparisons for the method factor for sample size $S$=200, 300, and 400.}
\label{tab:posthoc_anova_3way}
\begin{subtable}{\columnwidth}
\centering
\caption{Comparison among the proposed CBF features.}
\begin{tabular}{lccc}
\hline
\textbf{Comparison} & \textbf{Mean Difference} & \textbf{Std. Error} & \textbf{Adj. $p$-value} \\
\hline
CBF-$PC^+$ vs. CBF-$PC$    & 0.19524 & 0.00956 & $1.63\times10^{-53*}$ \\
CBF-$PC^+$ vs. CBF-$DC^+$  & 0.22078 & 0.00956 & $2.30\times10^{-58*}$ \\
CBF-$PC^+$ vs. CBF-$DC$    & 0.33360 & 0.00956 & $3.83\times10^{-75*}$ \\
CBF-$PC$ vs. CBF-$DC^+$    & 0.02554 & 0.00956 & $1.03\times10^{-2*}$ \\
CBF-$PC$ vs. CBF-$DC$      & 0.13836 & 0.00956 & $1.66\times10^{-40*}$ \\
CBF-$DC^+$ vs. CBF-$DC$    & 0.11282 & 0.00956 & $2.26\times10^{-33*}$ \\
\hline
\end{tabular}
\end{subtable}
\vspace{3mm}
\begin{subtable}{\columnwidth}
\centering
\caption{Comparison of the best-performing CBF feature (CBF-$PC^+$) with baseline methods.}
\begin{tabular}{lccc}
\hline
\textbf{Comparison} & \textbf{Mean Difference} & \textbf{Std. Error} & \textbf{Adj. $p$-value} \\
\hline
CBF-$PC^+$ vs. Corr-FC & 0.37202 & 0.00956 & $1.17\times10^{-79*}$ \\
CBF-$PC^+$ vs. GEFF    & 0.38758 & 0.00956 & $2.31\times10^{-81*}$ \\
CBF-$PC^+$ vs. NN      & 0.32862 & 0.00956 & $1.60\times10^{-74*}$ \\
CBF-$PC^+$ vs. SVM     & 0.69643 & 0.00956 & $5.58\times10^{-106*}$ \\
\hline
\end{tabular}
\end{subtable}
\end{table}

\subsubsection{Discriminative Connectivity Maps}
\label{DCM_results}
We characterize the spatial distribution of the most discriminative communities using the C-ICC framework described in Section~\ref{C-ICC}. Specifically, for each top-ranked partition, the detected communities are ranked according to their distinctiveness scores and mapped to DCMs. This analysis is repeated for the top-10 ranked partitions, and the resulting DCMs are averaged to obtain a summary representation. A higher percentage indicates that the corresponding brain system appears more frequently within these communities across subjects and across multiple joint community structure scales, i.e., different combinations of the resolution parameter $\gamma$ and inter-layer coupling parameter $\omega$.

Fig.~\ref{fig:Top1_400} presents the DCMs for all eight fMRI tasks, computed using the top-10 CBF-$PC^+$ fingerprints for sample size of $S=400$. Each DCM reflects the average contribution of the seven Yeo functional brain systems appearing within the top-1\% most discriminative communities across subjects. The corresponding DCMs for the top-5\% most discriminative communities are provided in the Supplementary Material. A biological interpretation of these findings is presented in Section~\ref{DCM_Dis}.

\subsection{Between-task fingerprinting}
\label{between_task}
For the between-task fingerprinting experiment, we follow the same procedure described in Section~\ref{within_task}, Steps (a)--(c), while modifying the construction of the training and testing CBFs in Step (d). Specifically, the selected top-$L$ ($L=10$) partitions are used to construct the training CBFs from the RL scans of a given task, whereas the testing CBFs are constructed from the unseen RL scans of each of the remaining task conditions. Subject identity is then determined using the distance-based classifier defined in Section~\ref{Sub_iden}.
This represents a substantially more challenging identification problem, as both the cognitive state and the underlying functional connectivity patterns differ between the training and testing scans. For every pair of tasks, the identification success rate was computed and summarized by an $8 \times 8$ confusion matrix.
To evaluate the scalability of the proposed framework, we performed large-scale experiments using sample sizes of $S=200$, 300, and 400. For each sample size, a different randomly selected group of subjects was used to construct the task-specific fingerprint databases and evaluate between-task identification. 

As shown in Fig. \ref{fig:between_400} (sample size $S=400$), the proposed framework consistently preserves subject identity across different task conditions despite the considerable task-induced reconfiguration of functional brain networks. Among the four proposed fingerprint representations, CBF-$PC^{+}$ achieves the strongest cross-task performance, producing the highest off-diagonal accuracies across nearly all task pairs. The remaining CBF features also substantially outperform the competing approaches, including Corr-FC, GEFF, SVM, and NN. In contrast, the baseline methods exhibit lower off-diagonal accuracies, indicating limited generalization when the training and testing scans originate from different cognitive states. Results for $S=200$ and $S=300$, which further support these findings, are provided in the Supplementary Material.

\vspace{-0.1in}

\section{Discussion}
Recent studies demonstrate that functional connectomes contain subject-specific \textit{fingerprints} capable of identifying individuals across sessions and tasks \cite{finn2015functional,gratton2018functional}. However, the topological basis of these fingerprints remains poorly understood. Characterizing individualized topological features enables defining normative patterns in healthy brains and quantifying subject-level deviations, shifting from group-based to personalized mapping. This approach has important clinical implications. For example, atypical connectivity in autism and disrupted communication in schizophrenia are associated with cognitive and behavioral impairments, highlighting the potential of individualized network fingerprints for understanding neuropsychiatric disorders \cite{monk2009abnormalities, sheffield2016cognition}.

Current fingerprinting methods rely primarily on edge-level connectivity and neglect higher-order network structure, limiting their interpretability \cite{finn2015functional,amico2018quest,li2021feature,abbas2020geff,mantwill2022brain}. We address this gap by introducing community-based functional connectome fingerprints that capture individual differences in mesoscale network organization. Rather than using traditional connectivity profiles as fingerprints, we represent each subject by a low-dimensional community-based fingerprint that captures the unique organization of the cortical system across fMRI conditions.
The proposed approach provides insights into how individual-specific community configurations vary across subjects, which may be valuable for studying personalized brain organization and its alterations in clinical populations.

\subsection{Within- and Between-task Fingerprinting Analysis}
\label{FA_Dis}
Across both within- and between-task settings, the proposed method outperforms state-of-the-art approaches. Corr-FC shows reduced performance in task fMRI due to its reliance on potentially noisy whole-network FC \cite{finn2015functional}, while GEFF exhibits sensitivity to preprocessing, sample size, and eigenvector selection. 
On the other hand, machine learning-based methods such as SVM and NN require sufficient training samples to learn robust subject-specific representations \cite{hannum2023high}. To ensure a fair comparison, all methods were evaluated under the same experimental settings, where only a single scan per subject was available for each condition. Under these constraints, the limited amount of training data reduces the ability of machine learning models to learn discriminative subject-specific patterns, resulting in lower identification accuracy. For this reason, deep learning-based fingerprinting methods are not suitable for comparison with our method for the current experimental settings \cite{hannum2023high,cai2021functional,griffa2022brain,lee2024discovering,lu2024brain}.
In contrast, by operating at the community level and integrating multiscale topological information, CBF yields more stable, interpretable, and less task-dependent fingerprints, enabling between-task identification with moderate accuracy. 

\subsection{Discriminative Connectivity Maps Interpretability} \label{DCM_Dis}
We demonstrated the utility of our method on a large-scale dataset of healthy individuals from the HCP study, successfully identifying meaningful discriminative connectivity maps. Although all tasks exhibit distributed contributions from multiple systems, the dominant networks differ across cognitive functions, demonstrating that subject-discriminative connectivity patterns are organized in a task-specific manner, as shown in Fig. ~\ref{fig:Top1_400}.

For the resting-state condition (Fig. ~\ref{Rest_DCM}), the default mode network (DMN) exhibits the largest contribution (49.4\%), followed by the frontoparietal network (FPN) (15.5\%) and visual (18.9\%) systems. This finding is consistent with the established role of the DMN in intrinsic brain organization and spontaneous cognition, suggesting that individual variability during rest is primarily encoded within internally oriented functional networks rather than sensory systems \cite{finn2015functional,amico2018quest, barch2013function}.

The working memory task (Fig. ~\ref{WM_DCM}) is characterized by a strong predominance of the dorsal attention network (DAN) (36.2\%), together with substantial contributions from the visual (22.7\%) and DMN (16.8\%) systems. This spatial organization agrees with the sustained attentional control and visuo-spatial processing required during working-memory performance, while the residual DMN contribution indicates that individual fingerprint information remains distributed across multiple large-scale networks.

For the motor task (Fig. ~\ref{MOT_DCM}), the somatosensory (SOM) (18.2\%) and FPN (28.6\%) systems contribute most strongly, accompanied by the DAN (20.9\%) and visual (13.5\%) networks. These results reflect the integration of sensorimotor execution with attentional and executive control processes required for coordinated movement, indicating that subject-specific motor fingerprints extend beyond primary motor cortices.

The social cognition task (Fig. ~\ref{SOC_DCM})  demonstrates the most pronounced network specialization, with nearly 80\% of the discriminative information localized within the DMN (79.8\%). This observation is consistent with the well-established involvement of medial prefrontal, posterior cingulate, and temporoparietal regions in mentalizing and social cognition, suggesting that inter-subject variability during social processing is dominated by DMN connectivity.

The language task (Fig. ~\ref{LAN_DCM}) similarly exhibits a dominant DMN contribution (45.7\%), together with limbic (12.8\%), ventral attention network (VAN) (11.4\%), and visual (10.4\%) contributions. Rather than reflecting isolated language areas, these findings indicate that individual differences in language-related functional organization arise from interactions among multiple association networks supporting semantic processing, attention, and memory.

For the gambling task (Fig. ~\ref{GAM_DCM}) , discriminative information is distributed across the SOM (27.7\%), DMN (18.6\%), visual (18.5\%), ventral attention (17.4\%), and limbic (13.6\%) systems. This widespread organization reflects the combined influence of sensory processing, reward evaluation, attentional orienting, and decision-making mechanisms that jointly contribute to individual variability during reward-guided behavior.

The emotion task (Fig. ~\ref{EMO_DCM}) is primarily represented by the DMN (30.9\%) and FPN (26.8\%) systems, with additional contributions from the Ventral Attention (15.3\%) and Somatosensory (10.6\%) networks. Although the limbic system contributes to the discriminative communities (6.0\%), it is not the dominant source of fingerprint information, suggesting that individual differences in emotional processing emerge from distributed interactions among association and attentional networks rather than limbic regions alone.

Finally, the relational reasoning task (Fig. ~\ref{REL_DCM}) exhibits the strongest contributions from the DMN (31.6\%) and SOM (30.4\%) systems, followed by the limbic (15.6\%) and FPN (12.3\%) networks. This distributed pattern is consistent with the integration of executive reasoning, memory retrieval, and sensory representations required during higher-order relational processing.

The DCMs demonstrate that subject-specific fingerprints are neither globally distributed nor confined to a single functional network. Instead, each cognitive state is characterized by a unique spatial profile of discriminative connectivity, with higher-order association networks, including the Default Mode Network, Frontoparietal Network, and Dorsal Attention Network, consistently contributing across tasks. These findings suggest that individual variability is primarily encoded within large-scale integrative networks. The consistent emergence of well-established cognitive systems within the most discriminative communities supports the biological plausibility of the proposed framework, indicating that the derived fingerprints capture meaningful large-scale functional brain organization rather than random regional fluctuations \cite{finn2015functional,betzel2019community,barch2013function}.

Comparing the top-1\% and top-5\% DCMs (provided in the Supplementary Material) reveals a high degree of spatial consistency across all eight tasks. The dominant functional systems identified in the top-1\% analysis remain dominant in the top-5\% DCMs, demonstrating that the proposed framework captures robust and reproducible task-specific fingerprint patterns rather than relying on a few isolated communities. The overall task-specific spatial signatures are largely preserved, indicating that discriminative connectivity is organized hierarchically, where the highest-ranked communities capture the core fingerprint while additional highly discriminative communities refine this representation without fundamentally altering its functional organization.

\subsection{Limitations and Future Work}
The primary limitation of this study is the scalability of modularity maximization for larger subject cohorts. A related challenge is the variability inherent to community detection, arising from the degeneracy of the modularity landscape and the stochastic nature of greedy optimization  \cite{blondel2008fast}. To address these limitations, future work will incorporate parallel and distributed-memory implementations of the Louvain algorithm to enable simultaneous evaluation of modularity gains across nodes, thereby improving computational efficiency \cite{ghosh2018distributed,wang2025swift,mohammadi2021accelerating}. In addition, we will explore different clustering strategies, non-random initialization schemes, and degree-based node ordering to mitigate stochastic variability in modularity optimization \cite{aviyente2022explainability}.

Future work will also consider extension to dynamic functional connectivity (dFC) to examine whether the derived fingerprints preserve their stability and individuality across temporal fluctuations in brain connectivity. Finally, integrating CBF-derived fingerprints with cognitive and behavioral measures may provide deeper insight into the biological and phenotypic basis of individual variability, further enhancing the interpretability and potential clinical relevance of the proposed approach.

\vspace{-0.1in}

\section{Conclusion}
This paper introduces a new perspective on brain fingerprinting by emphasizing individual differences in the mesoscale organization of functional networks. 
Instead of relying on traditional connectivity profiles, each subject is represented by a community-based fingerprint that reflects how cortical systems organize across fMRI tasks and sessions. We propose a signed multilayer community detection framework that incorporates both correlations and anticorrelations to identify subject-specific communities and derive graph-theoretic fingerprints capturing within-task structure and between-task reconfiguration. The proposed approach produces stable, interpretable, and highly individualized representations of brain function, supporting its potential for precision neuroimaging and personalized neuroscience.

\vspace{-0.1in}
\section*{Acknowledgment}
This work was supported in part by the Air Force Office of Scientific Research Grant FA9550-23-1-0224. 
\vspace{-0.1in}
\section{Compliance with Ethical Standards}
This research study was conducted retrospectively using human subject data made available in open access by \cite{gordon2017}. Ethical approval was not required as confirmed by the license attached with the open-access data.

\bibliographystyle{unsrt}
\bibliography{refs-sema}

\begin{thebibliography}{10}

\bibitem{biswal1995functional}
Bharat Biswal, F~Zerrin~Yetkin, Victor~M Haughton, and James~S Hyde.
\newblock Functional connectivity in the motor cortex of resting human brain using echo-planar mri.
\newblock {\em Magnetic resonance in medicine}, 34(4):537--541, 1995.

\bibitem{finn2015functional}
Emily~S Finn, Xilin Shen, Dustin Scheinost, Monica~D Rosenberg, Jessica Huang, Marvin~M Chun, Xenophon Papademetris, and R~Todd Constable.
\newblock Functional connectome fingerprinting: identifying individuals using patterns of brain connectivity.
\newblock {\em Nature neuroscience}, 18(11):1664--1671, 2015.

\bibitem{gratton2018functional}
Caterina Gratton, Timothy~O Laumann, Ashley~N Nielsen, Deanna~J Greene, Evan~M Gordon, Adrian~W Gilmore, Steven~M Nelson, Rebecca~S Coalson, Abraham~Z Snyder, Bradley~L Schlaggar, et~al.
\newblock Functional brain networks are dominated by stable group and individual factors, not cognitive or daily variation.
\newblock {\em Neuron}, 98(2):439--452, 2018.

\bibitem{finn2021beyond}
Emily~S Finn and Monica~D Rosenberg.
\newblock Beyond fingerprinting: Choosing predictive connectomes over reliable connectomes.
\newblock {\em NeuroImage}, 239:118254, 2021.

\bibitem{amico2018quest}
Enrico Amico and Joaqu{\'\i}n Go{\~n}i.
\newblock The quest for identifiability in human functional connectomes.
\newblock {\em Scientific reports}, 8(1):8254, 2018.

\bibitem{pena2018spatiotemporal}
Cleof{\'e} Pe{\~n}a-G{\'o}mez, Andrea Avena-Koenigsberger, Jorge Sepulcre, and Olaf Sporns.
\newblock Spatiotemporal network markers of individual variability in the human functional connectome.
\newblock {\em Cerebral Cortex}, 28(8):2922--2934, 2018.

\bibitem{abbas2020geff}
Kausar Abbas, Enrico Amico, Diana~Otero Svaldi, Uttara Tipnis, Duy~Anh Duong-Tran, Mintao Liu, Meenusree Rajapandian, Jaroslaw Harezlak, Beau~M Ances, and Joaqu{\'\i}n Go{\~n}i.
\newblock Geff: Graph embedding for functional fingerprinting.
\newblock {\em NeuroImage}, 221:117181, 2020.

\bibitem{carvalho2025functional}
Vitor Carvalho, Mintao Liu, Jaroslaw Harezlak, Ana~Mar{\'\i}a Estrada~G{\'o}mez, and Joaqu{\'\i}n Go{\~n}i.
\newblock Functional connectome fingerprinting through tucker tensor decomposition.
\newblock {\em Applied Sciences}, 15(9):4821, 2025.

\bibitem{hannum2023high}
Andrew Hannum, Mario~A Lopez, Sa{\'u}l~A Blanco, and Richard~F Betzel.
\newblock High-accuracy machine learning techniques for functional connectome fingerprinting and cognitive state decoding.
\newblock {\em Human Brain Mapping}, 44(16):5294--5308, 2023.

\bibitem{cai2021functional}
Biao Cai, Gemeng Zhang, Aiying Zhang, Li~Xiao, Wenxing Hu, Julia~M Stephen, Tony~W Wilson, Vince~D Calhoun, and Yu-Ping Wang.
\newblock Functional connectome fingerprinting: identifying individuals and predicting cognitive functions via autoencoder.
\newblock {\em Human Brain Mapping}, 42(9):2691--2705, 2021.

\bibitem{griffa2022brain}
Alessandra Griffa, Enrico Amico, Rapha{\"e}l Li{\'e}geois, Dimitri Van De~Ville, and Maria~Giulia Preti.
\newblock Brain structure-function coupling provides signatures for task decoding and individual fingerprinting.
\newblock {\em NeuroImage}, 250:118970, 2022.

\bibitem{lee2024discovering}
Juhyeon Lee and Jong-Hwan Lee.
\newblock Discovering individual fingerprints in resting-state functional connectivity using deep neural networks.
\newblock {\em Human Brain Mapping}, 45(1):e26561, 2024.

\bibitem{lu2024brain}
Jiayu Lu, Tianyi Yan, Lan Yang, Xi~Zhang, Jiaxin Li, Dandan Li, Jie Xiang, and Bin Wang.
\newblock Brain fingerprinting and cognitive behavior predicting using functional connectome of high inter-subject variability.
\newblock {\em NeuroImage}, 295:120651, 2024.

\bibitem{st2023functional}
Fr{\'e}d{\'e}ric St-Onge, Mohammadali Javanray, Alexa Pichet~Binette, Cherie Strikwerda-Brown, Jordana Remz, R~Nathan Spreng, Golia Shafiei, Bratislav Misic, {\'E}tienne Vachon-Presseau, and Sylvia Villeneuve.
\newblock Functional connectome fingerprinting across the lifespan.
\newblock {\em Network Neuroscience}, 7(3):1206--1227, 2023.

\bibitem{sporns2016modular}
Olaf Sporns and Richard~F Betzel.
\newblock Modular brain networks.
\newblock {\em Annual review of psychology}, 67(1):613--640, 2016.

\bibitem{Mucha2010}
Peter~J. Mucha, Thomas Richardson, Kevin Macon, Mason~A. Porter, and Jukka-Pekka Onnela.
\newblock Community structure in time-dependent, multiscale, and multiplex networks.
\newblock {\em Science}, 328(5980):876--878, 2010.

\bibitem{betzel2017multi}
Richard~F Betzel and Danielle~S Bassett.
\newblock Multi-scale brain networks.
\newblock {\em Neuroimage}, 160:73--83, 2017.

\bibitem{article}
Mark Newman and Michelle Girvan.
\newblock Finding and evaluating community structure in networks.
\newblock {\em Physical review. E, Statistical, nonlinear, and soft matter physics}, 69:026113, 03 2004.

\bibitem{Fortunato2016}
Santo Fortunato and Darko Hric.
\newblock Community detection in networks: A user guide.
\newblock {\em Physics Reports}, 659:1--44, 2016.
\newblock Community detection in networks: A user guide.

\bibitem{davis1967clustering}
James~A Davis.
\newblock Clustering and structural balance in graphs.
\newblock {\em Human relations}, 20(2):181--187, 1967.

\bibitem{gomez2009analysis}
Sergio G{\'o}mez, Pablo Jensen, and Alex Arenas.
\newblock Analysis of community structure in networks of correlated data.
\newblock {\em Physical Review E—Statistical, Nonlinear, and Soft Matter Physics}, 80(1):016114, 2009.

\bibitem{betzel2019community}
Richard~F Betzel, Maxwell~A Bertolero, Evan~M Gordon, Caterina Gratton, Nico~UF Dosenbach, and Danielle~S Bassett.
\newblock The community structure of functional brain networks exhibits scale-specific patterns of inter-and intra-subject variability.
\newblock {\em Neuroimage}, 202:115990, 2019.

\bibitem{blondel2008fast}
Vincent~D Blondel, Jean-Loup Guillaume, Renaud Lambiotte, and Etienne Lefebvre.
\newblock Fast unfolding of communities in large networks.
\newblock {\em Journal of statistical mechanics: theory and experiment}, 2008(10):P10008, 2008.

\bibitem{garcia2018applications}
Javier~O Garcia, Arian Ashourvan, Sarah Muldoon, Jean~M Vettel, and Danielle~S Bassett.
\newblock Applications of community detection techniques to brain graphs: Algorithmic considerations and implications for neural function.
\newblock {\em Proceedings of the IEEE}, 106(5):846--867, 2018.

\bibitem{guimera2005cartography}
Roger Guimera and Lu{\'\i}s A~Nunes Amaral.
\newblock Cartography of complex networks: modules and universal roles.
\newblock {\em Journal of Statistical Mechanics: Theory and Experiment}, 2005(02):P02001, 2005.

\bibitem{rubinov2011weight}
Mikail Rubinov and Olaf Sporns.
\newblock Weight-conserving characterization of complex functional brain networks.
\newblock {\em Neuroimage}, 56(4):2068--2079, 2011.

\bibitem{koo2016guideline}
Terry~K Koo and Mae~Y Li.
\newblock A guideline of selecting and reporting intraclass correlation coefficients for reliability research.
\newblock {\em Journal of chiropractic medicine}, 15(2):155--163, 2016.

\bibitem{sareen2021exploring}
Ekansh Sareen, S{\'e}lima Zahar, Dimitri Van De~Ville, Anubha Gupta, Alessandra Griffa, and Enrico Amico.
\newblock Exploring meg brain fingerprints: Evaluation, pitfalls, and interpretations.
\newblock {\em NeuroImage}, 240:118331, 2021.

\bibitem{yeo2011organization}
BT~Thomas Yeo, Fenna~M Krienen, Jorge Sepulcre, Mert~R Sabuncu, Danial Lashkari, Marisa Hollinshead, Joshua~L Roffman, Jordan~W Smoller, Lilla Z{\"o}llei, Jonathan~R Polimeni, et~al.
\newblock The organization of the human cerebral cortex estimated by intrinsic functional connectivity.
\newblock {\em Journal of neurophysiology}, 2011.

\bibitem{gao2021smooth}
Siyuan Gao, Xinyue Xia, Dustin Scheinost, and Gal Mishne.
\newblock Smooth graph learning for functional connectivity estimation.
\newblock {\em NeuroImage}, 239:118289, 2021.

\bibitem{van2013wu}
David~C Van~Essen, Stephen~M Smith, Deanna~M Barch, Timothy~EJ Behrens, Essa Yacoub, Kamil Ugurbil, Wu-Minn~HCP Consortium, et~al.
\newblock The wu-minn human connectome project: an overview.
\newblock {\em Neuroimage}, 80:62--79, 2013.

\bibitem{glasser2016multi}
Matthew~F Glasser, Timothy~S Coalson, Emma~C Robinson, Carl~D Hacker, John Harwell, Essa Yacoub, Kamil Ugurbil, Jesper Andersson, Christian~F Beckmann, Mark Jenkinson, et~al.
\newblock A multi-modal parcellation of human cerebral cortex.
\newblock {\em Nature}, 536(7615):171--178, 2016.

\bibitem{geisser1958extension}
Seymour Geisser and Samuel~W Greenhouse.
\newblock An extension of box's results on the use of the f distribution in multivariate analysis.
\newblock {\em The Annals of Mathematical Statistics}, pages 885--891, 1958.

\bibitem{huynh1976estimation}
Huynh Huynh and Leonard~S Feldt.
\newblock Estimation of the box correction for degrees of freedom from sample data in randomized block and split-plot designs.
\newblock {\em Journal of educational statistics}, 1(1):69--82, 1976.

\bibitem{monk2009abnormalities}
Christopher~S Monk, Scott~J Peltier, Jillian~Lee Wiggins, Shih-Jen Weng, Melisa Carrasco, Susan Risi, and Catherine Lord.
\newblock Abnormalities of intrinsic functional connectivity in autism spectrum disorders.
\newblock {\em Neuroimage}, 47(2):764--772, 2009.

\bibitem{sheffield2016cognition}
Julia~M Sheffield and Deanna~M Barch.
\newblock Cognition and resting-state functional connectivity in schizophrenia.
\newblock {\em Neuroscience \& Biobehavioral Reviews}, 61:108--120, 2016.

\bibitem{li2021feature}
Kendrick Li, Krista Wisner, and Gowtham Atluri.
\newblock Feature selection framework for functional connectome fingerprinting.
\newblock {\em Human Brain Mapping}, 42(12):3717--3732, 2021.

\bibitem{mantwill2022brain}
Maron Mantwill, Martin Gell, Stephan Krohn, and Carsten Finke.
\newblock Brain connectivity fingerprinting and behavioural prediction rest on distinct functional systems of the human connectome.
\newblock {\em Communications biology}, 5(1):261, 2022.

\bibitem{barch2013function}
Deanna~M Barch, Gregory~C Burgess, Michael~P Harms, Steven~E Petersen, Bradley~L Schlaggar, Maurizio Corbetta, Matthew~F Glasser, Sandra Curtiss, Sachin Dixit, Cindy Feldt, et~al.
\newblock Function in the human connectome: task-fmri and individual differences in behavior.
\newblock {\em Neuroimage}, 80:169--189, 2013.

\bibitem{ghosh2018distributed}
Sayan Ghosh, Mahantesh Halappanavar, Antonino Tumeo, Ananth Kalyanaraman, Hao Lu, Daniel Chavarria-Miranda, Arif Khan, and Assefaw Gebremedhin.
\newblock Distributed louvain algorithm for graph community detection.
\newblock In {\em 2018 IEEE international parallel and distributed processing symposium (IPDPS)}, pages 885--895. IEEE, 2018.

\bibitem{wang2025swift}
Zhibin Wang, Xi~Lin, Xue Li, Pinhuan Wang, Ziheng Meng, Hang Liu, Chen Tian, and Sheng Zhong.
\newblock Swift unfolding of communities: Gpu-accelerated louvain algorithm.
\newblock In {\em Proceedings of the 30th ACM SIGPLAN Annual Symposium on Principles and Practice of Parallel Programming}, pages 441--454, 2025.

\bibitem{mohammadi2021accelerating}
Maryam Mohammadi, Mahmood Fazlali, and Mehdi Hosseinzadeh.
\newblock Accelerating louvain community detection algorithm on graphic processing unit.
\newblock {\em The Journal of supercomputing}, 77(6):6056--6077, 2021.

\bibitem{aviyente2022explainability}
Selin Aviyente and Abdullah Karaaslanli.
\newblock Explainability in graph data science: Interpretability, replicability, and reproducibility of community detection.
\newblock {\em IEEE Signal Processing Magazine}, 39(4):25--39, 2022.

\bibitem{gordon2017}
Evan~M Gordon, Timothy~O Laumann, Adrian~W Gilmore, Dillan~J Newbold, Deanna~J Greene, Jeffrey~J Berg, Mario Ortega, Catherine Hoyt-Drazen, Caterina Gratton, Haoxin Sun, et~al.
\newblock Precision functional mapping of individual human brains.
\newblock {\em Neuron}, 95(4):791--807, 2017.

\end{thebibliography}

\end{document}


\maketitle

\section{Sensitivity Analysis of Top-$L$ partitions}
\label{Supp_sec2}
The multi-scale partitions obtained by optimizing the signed multilayer modularity (SMM) objective on the multi-subject networks were used to extract community-based fingerprints (CBFs), yielding one CBF per subject for each partition. These CBFs were ranked according to their discriminative power using the mean pairwise Euclidean distance (MPED) criterion, with the top-$L$ ranked CBFs corresponding  to the top-$L$  partitions.

\begin{figure*}[t]
\centering
\begin{subfigure}{0.35\textwidth}
    \centering
    \includegraphics[width=\linewidth]{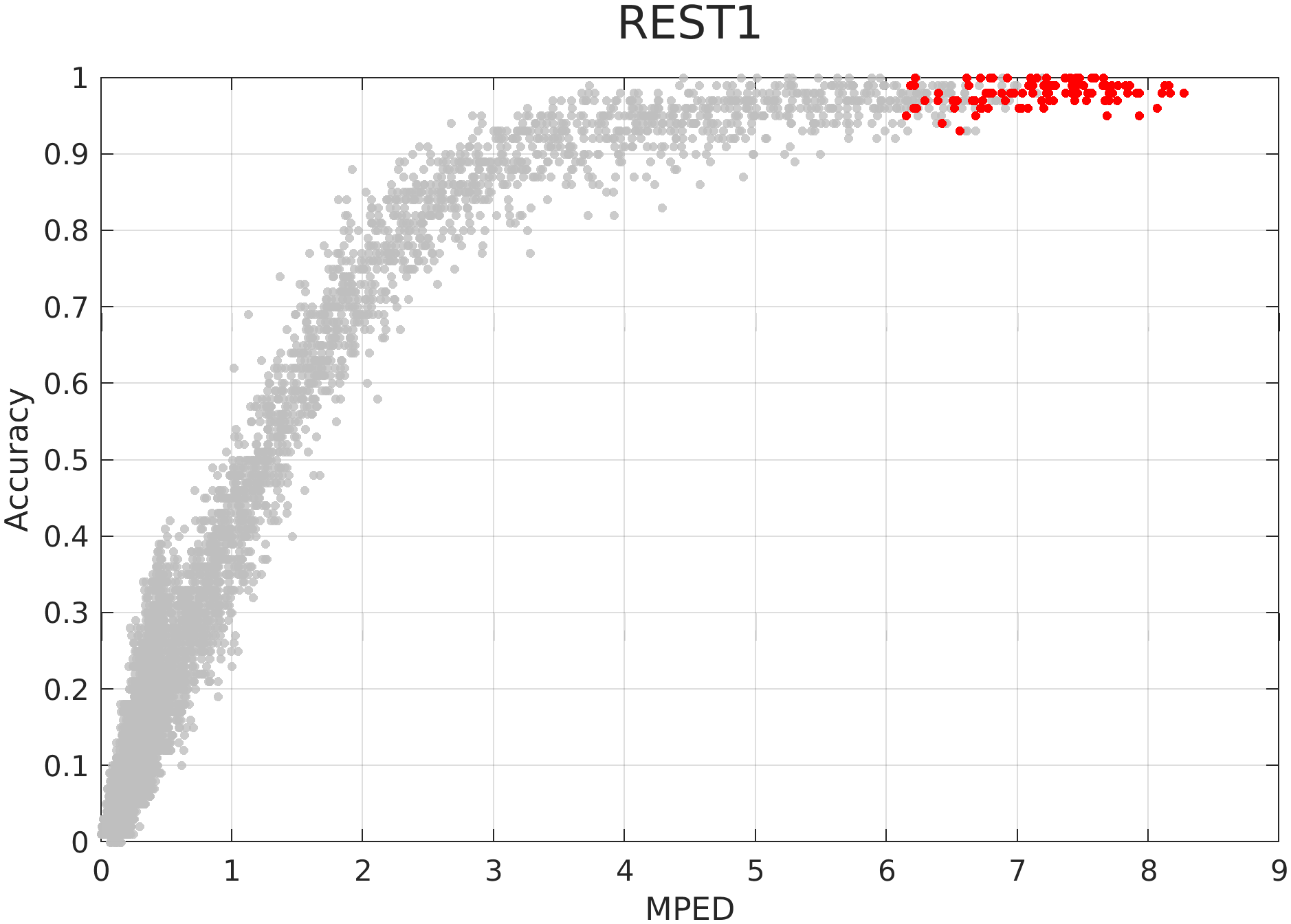}
    \caption{REST1}
\end{subfigure}
\hfill
\begin{subfigure}{0.35\textwidth}
    \centering
    \includegraphics[width=\linewidth]{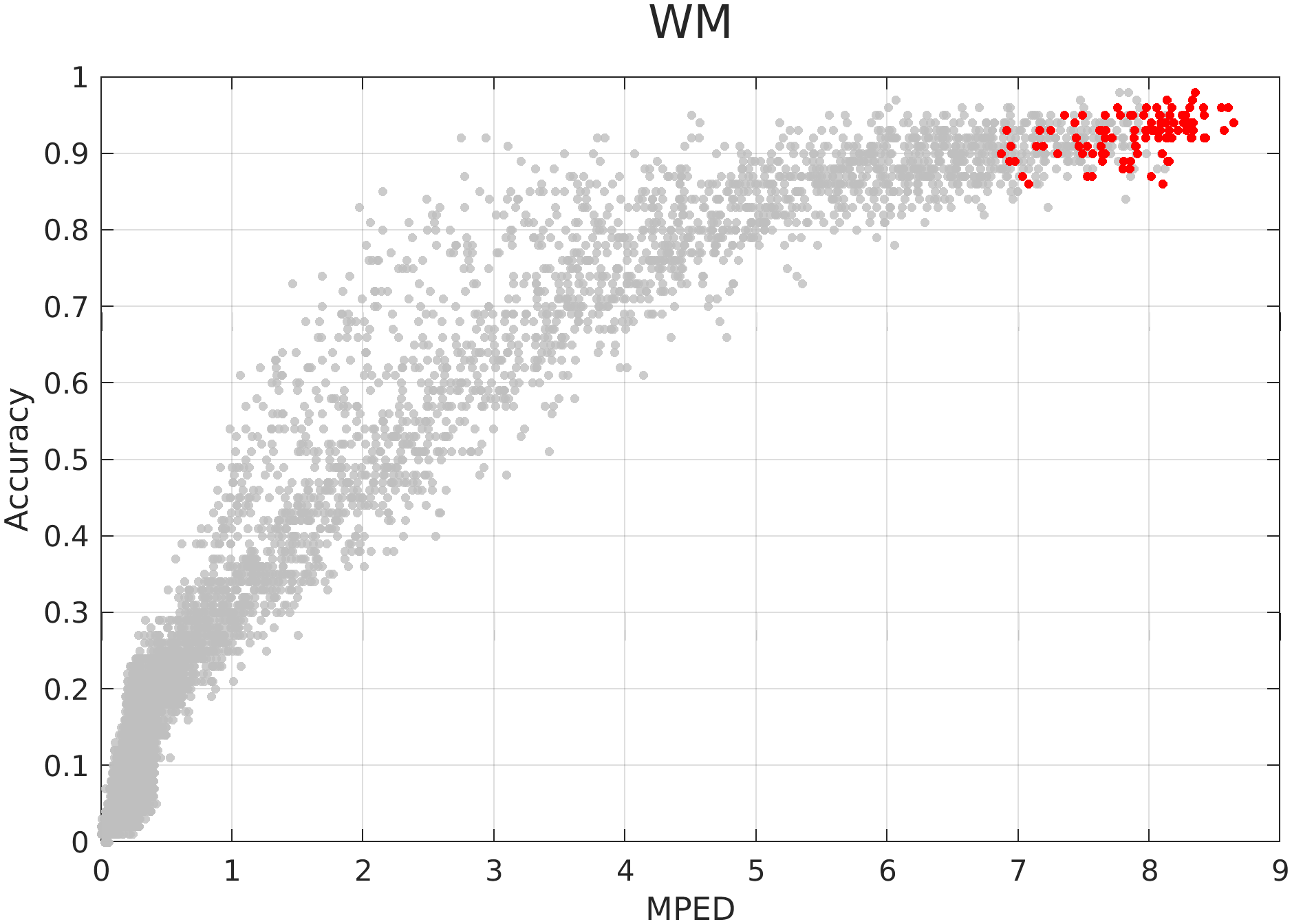}
    \caption{WM}
\end{subfigure}

\vspace{0.5em}

\begin{subfigure}{0.35\textwidth}
    \centering
    \includegraphics[width=\linewidth]{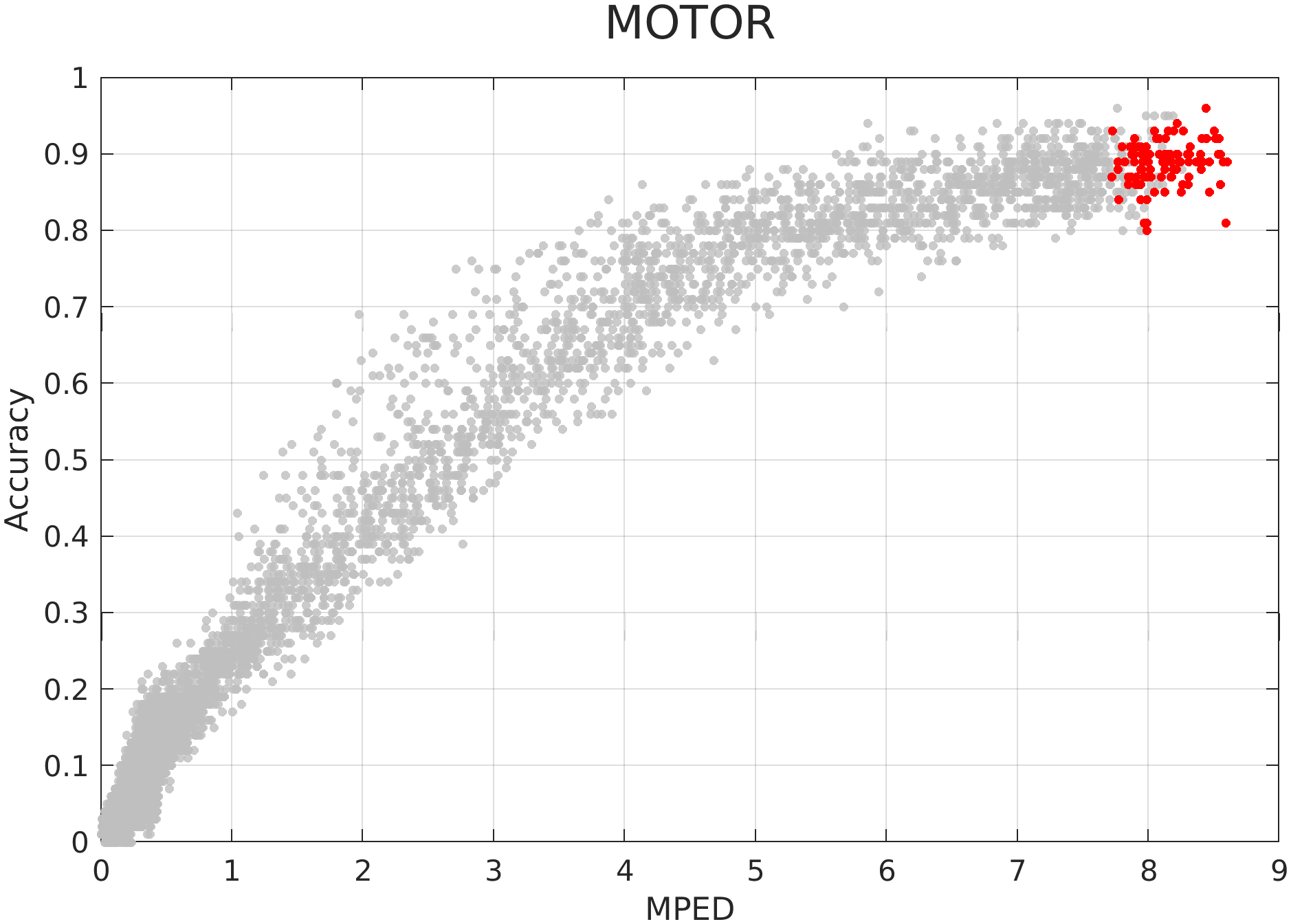}
    \caption{MOTOR}
\end{subfigure}
\hfill
\begin{subfigure}{0.35\textwidth}
    \centering
    \includegraphics[width=\linewidth]{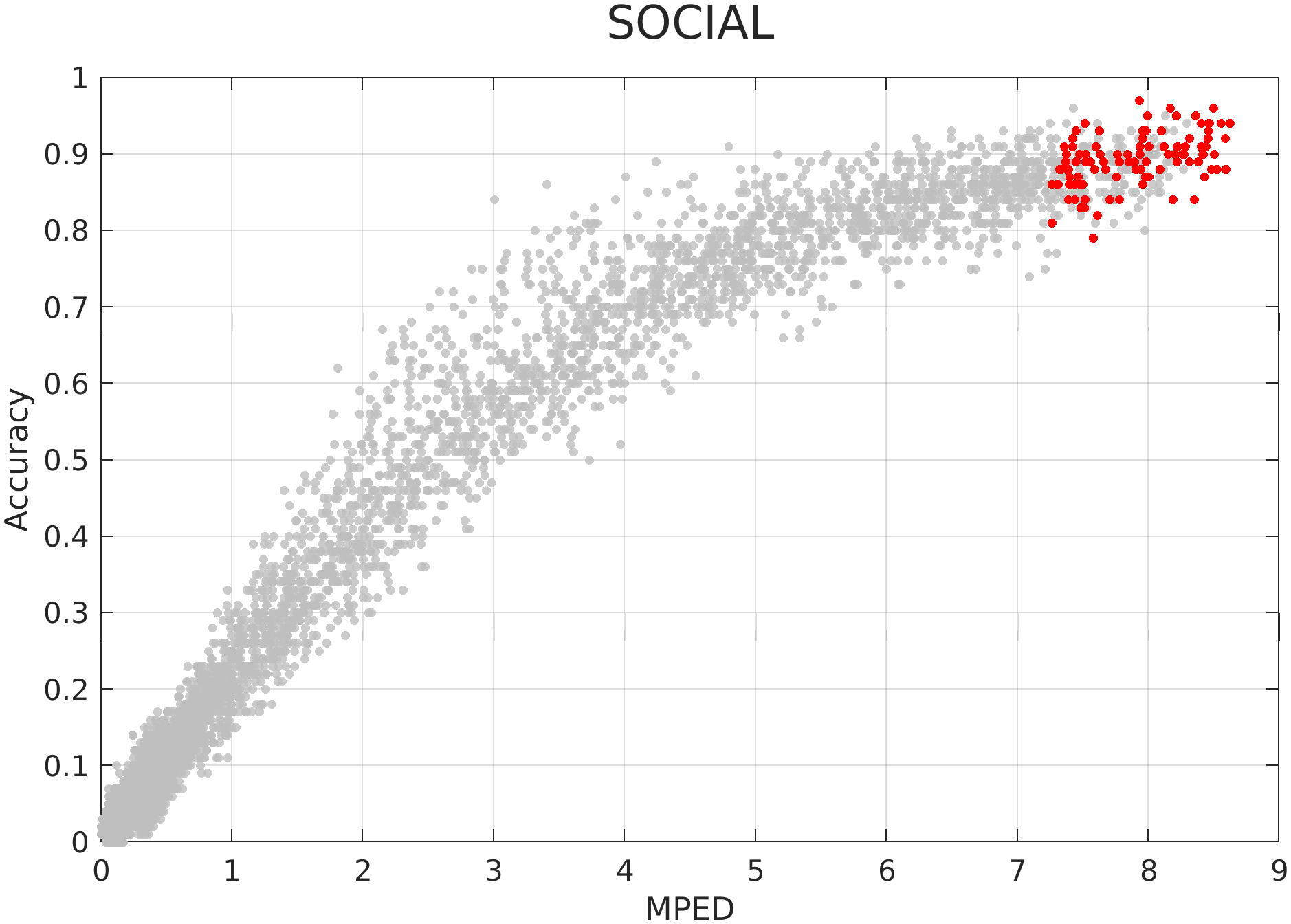}
    \caption{SOCIAL}
\end{subfigure}
\vspace{0.5em}
\begin{subfigure}{0.35\textwidth}
    \centering
    \includegraphics[width=\linewidth]{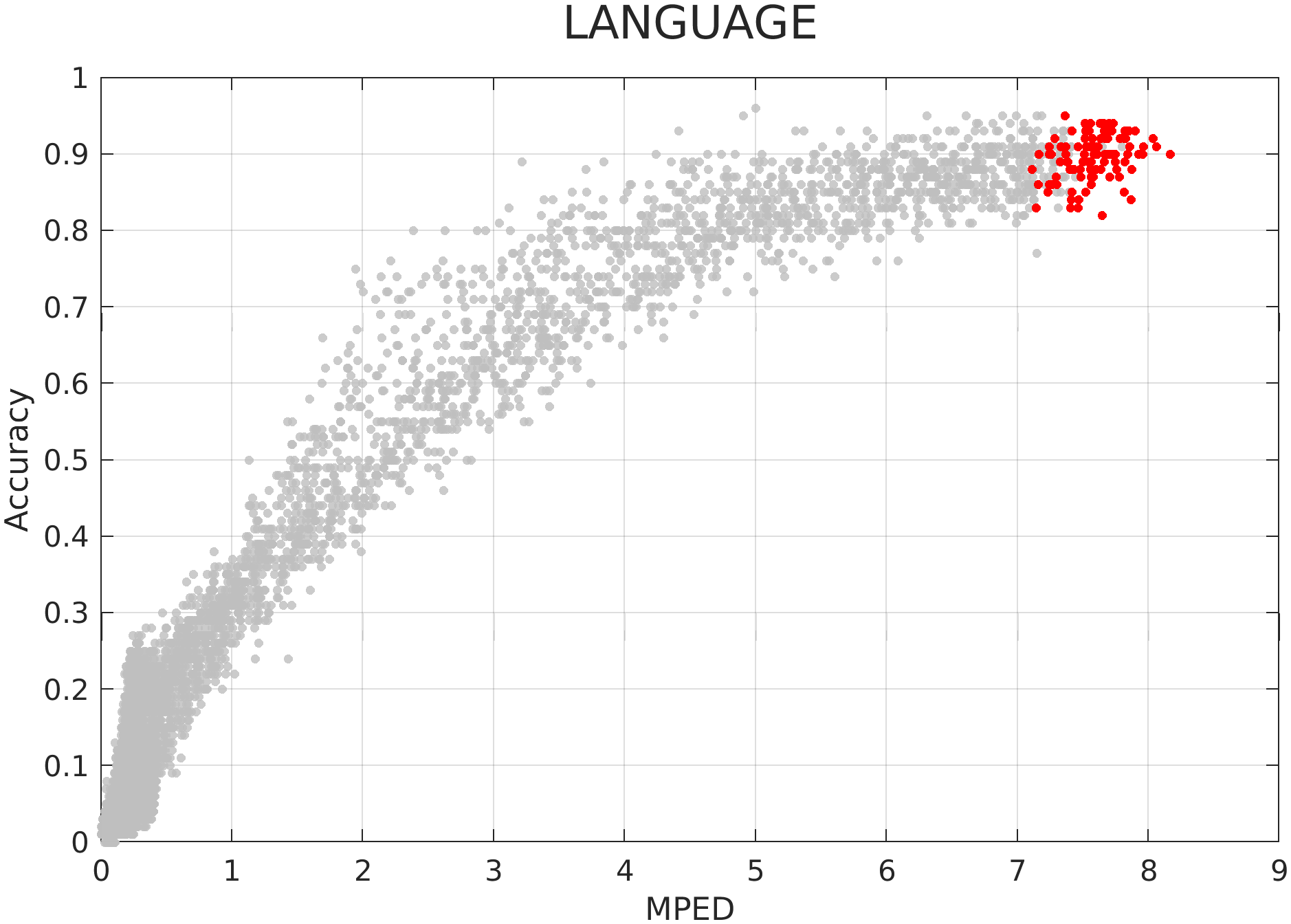}
    \caption{LANGUAGE}
\end{subfigure}
\hfill
\begin{subfigure}{0.35\textwidth}
    \centering
    \includegraphics[width=\linewidth]{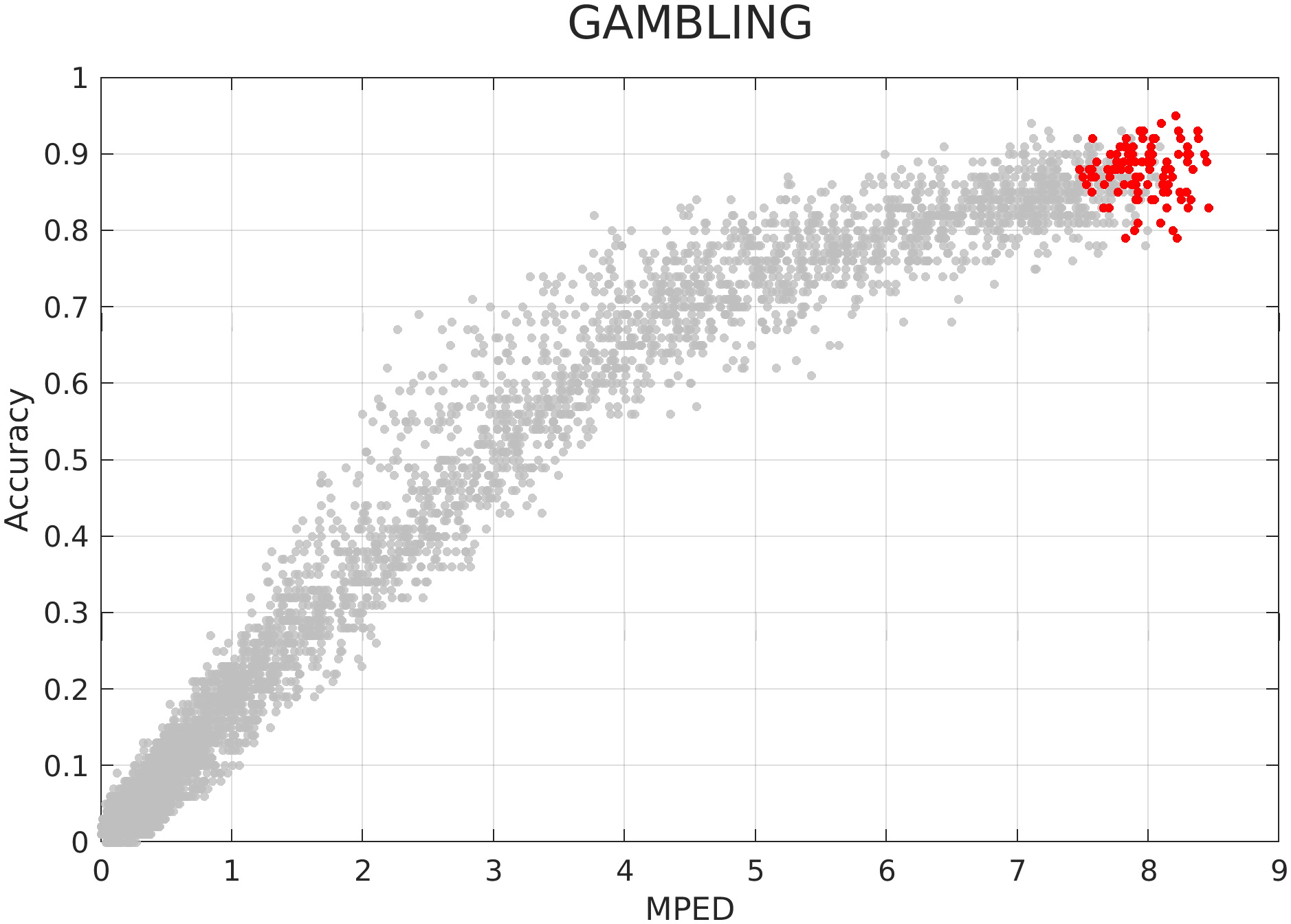}
    \caption{GAMBLING}
\end{subfigure}

\vspace{0.5em}

\begin{subfigure}{0.35\textwidth}
    \centering
    \includegraphics[width=\linewidth]{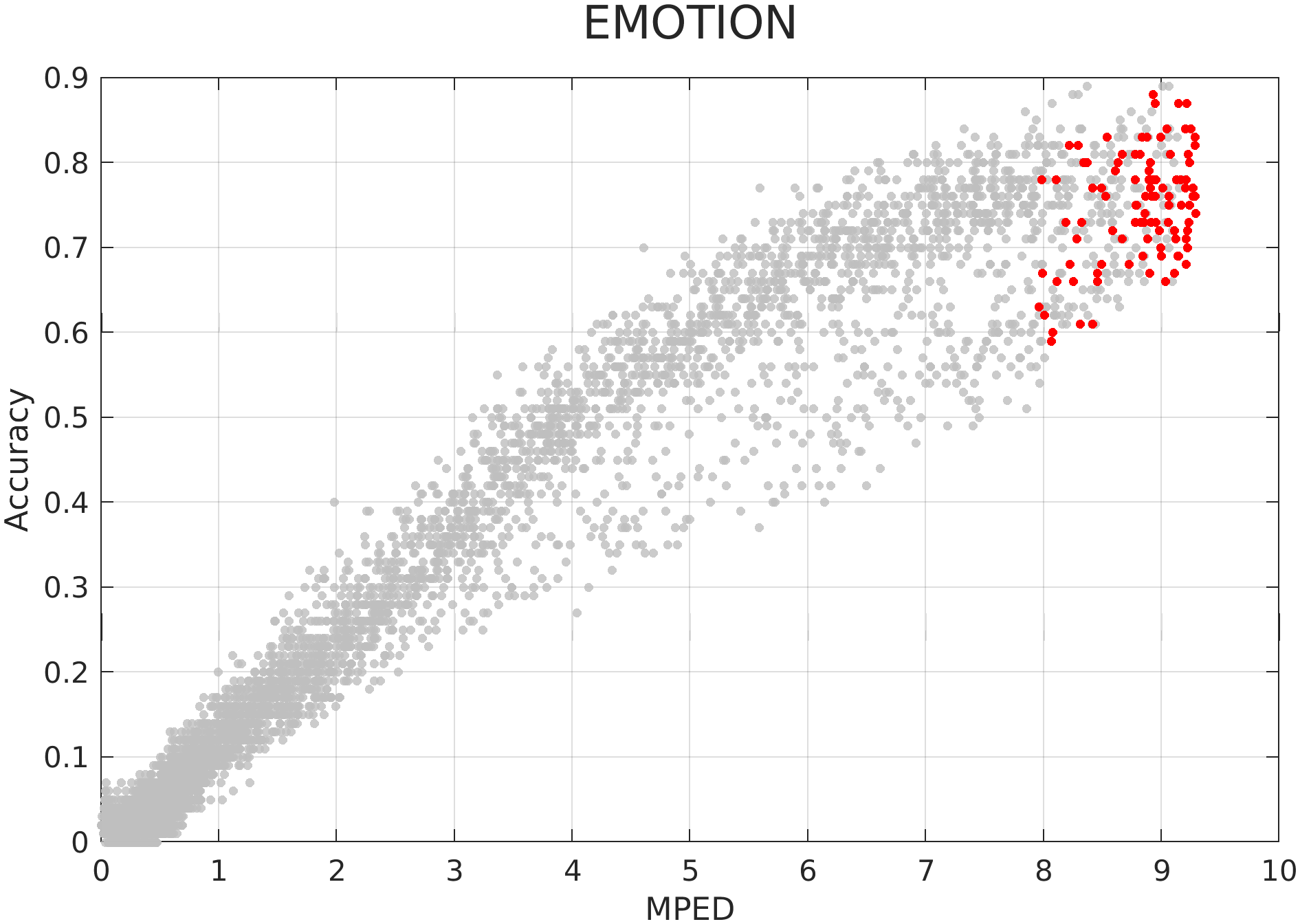}
    \caption{EMOTION}
\end{subfigure}
\hfill
\begin{subfigure}{0.35\textwidth}
    \centering
    \includegraphics[width=\linewidth]{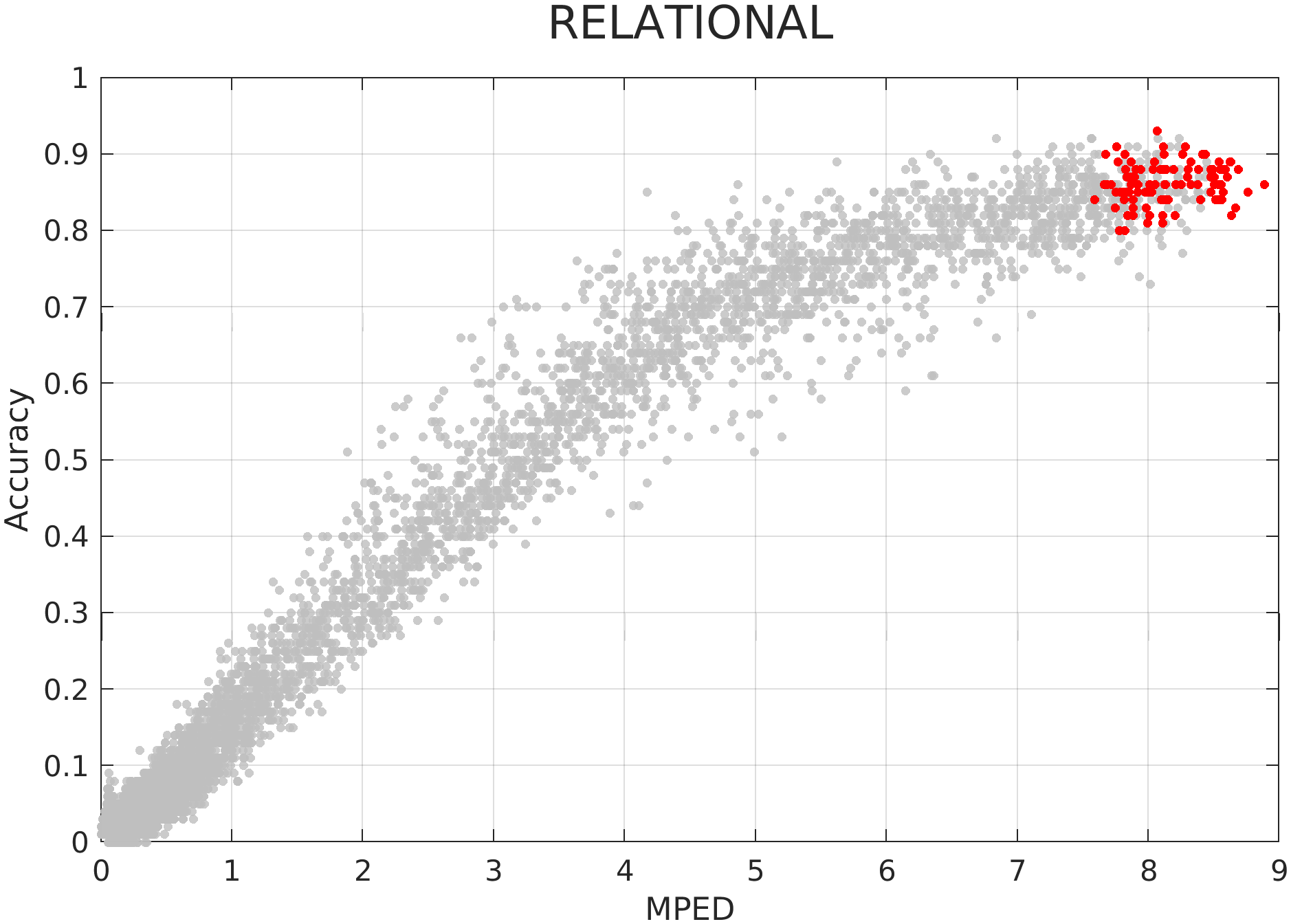}
    \caption{RELATIONAL}
\end{subfigure}
\caption{Relationship between the mean pairwise Euclidean distance (MPED) and identification accuracy for all generated partitions. Each point represents a partition and its corresponding MPED and identification accuracy values. Gray points denote all candidate partitions obtained across 10 independent runs, while red points indicate the top-10 partitions selected using the 99th-percentile MPED criterion. Each run was performed on a different randomly sampled cohort of subjects and evaluated independently; for visualization purposes, partitions from all runs are aggregated into a single scatter plot for each task.}
\label{fig:SA_S}
\end{figure*}

\begin{figure*}[t]
\centering
\begin{subfigure}{0.48\textwidth}
    \centering
    \includegraphics[width=\linewidth]{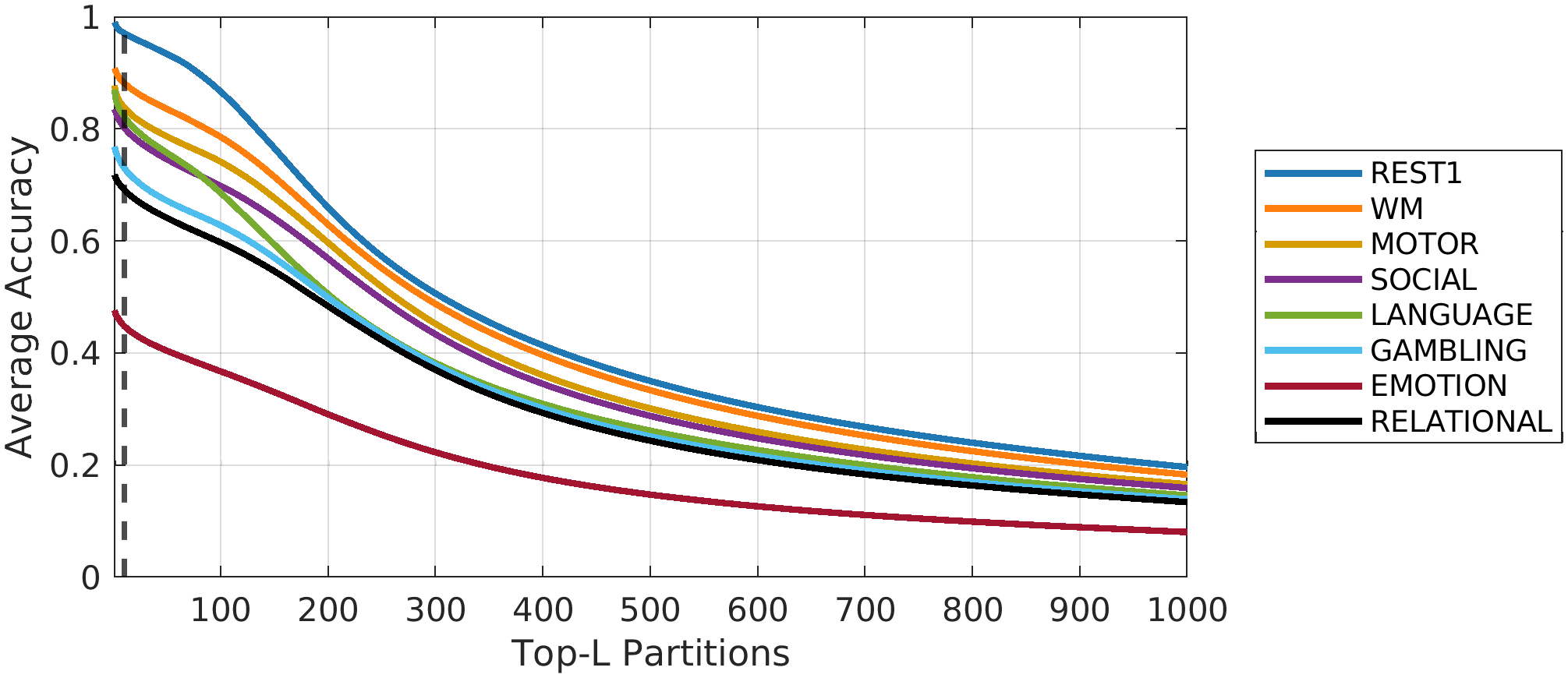}
    \caption{$PC$}
\end{subfigure}
\hfill
\begin{subfigure}{0.48\textwidth}
    \centering
    \includegraphics[width=\linewidth]{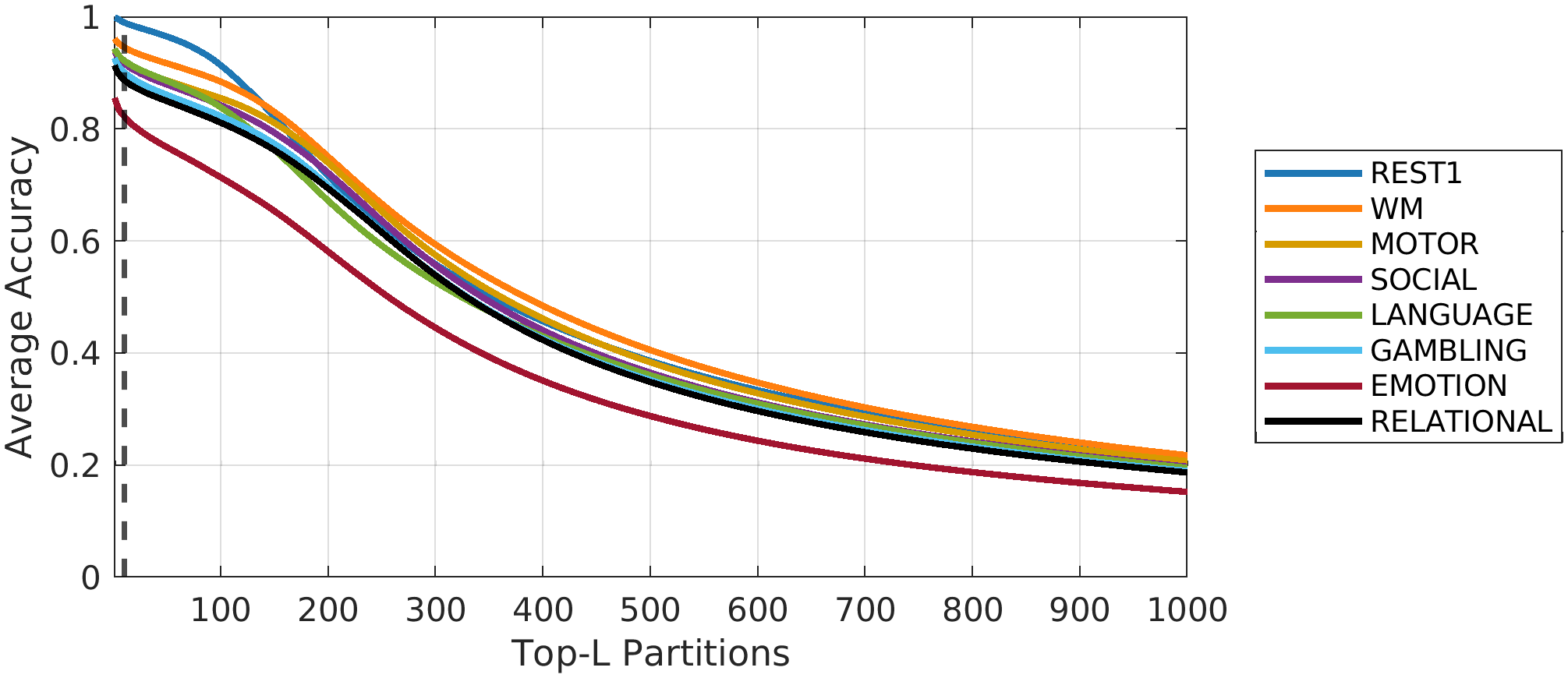}
    \caption{$PC^+$}
\end{subfigure}
\vspace{0.5em}
\begin{subfigure}{0.48\textwidth}
    \centering
    \includegraphics[width=\linewidth]{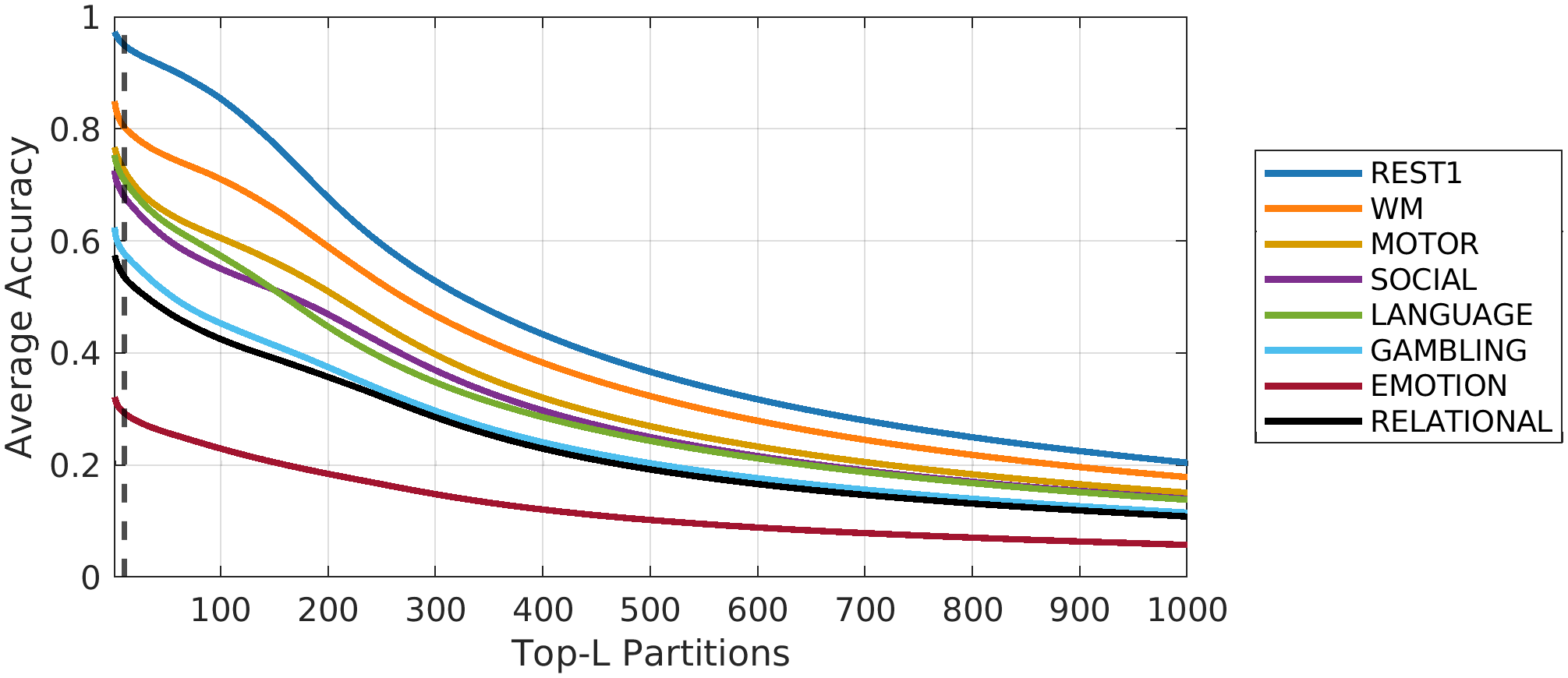}
    \caption{$DC$}
\end{subfigure}
\hfill
\begin{subfigure}{0.48\textwidth}
    \centering
    \includegraphics[width=\linewidth]{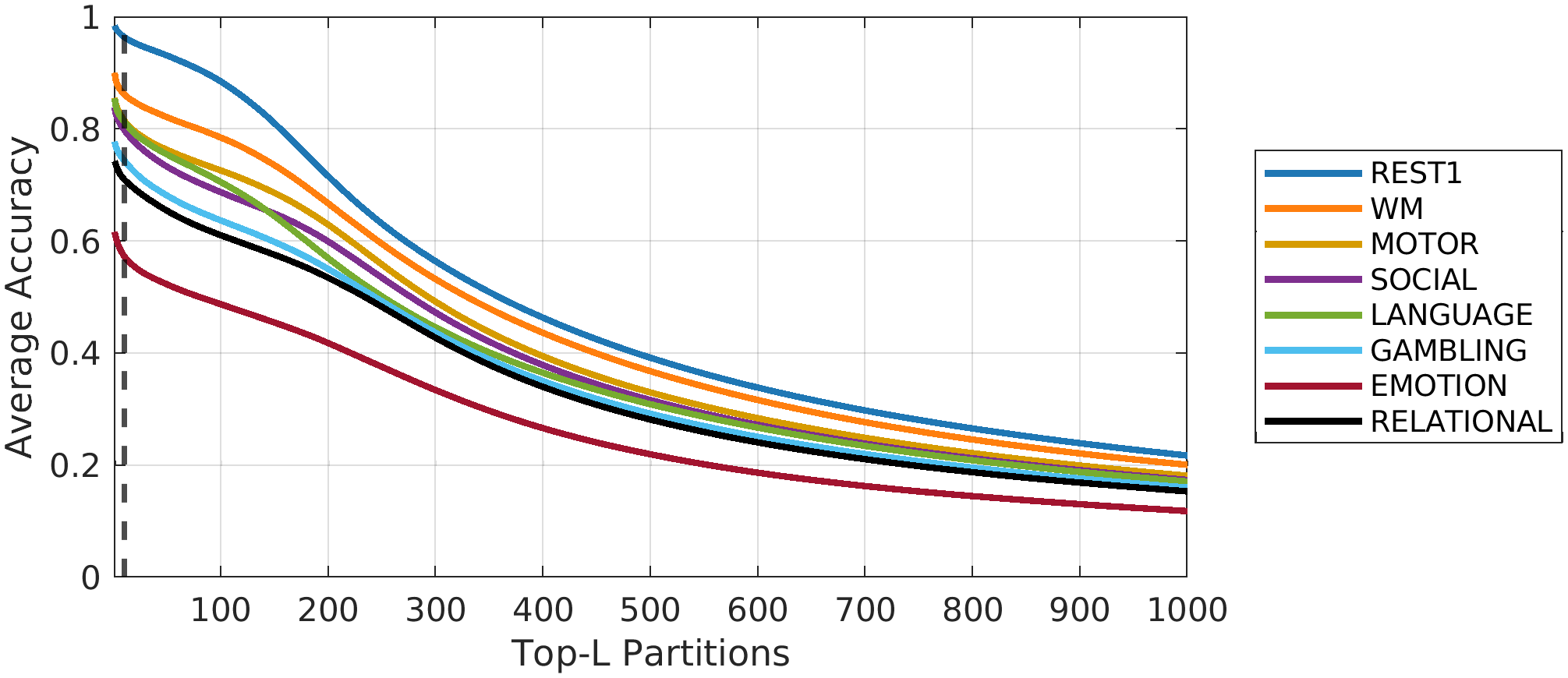}
    \caption{$DC^+$}
\end{subfigure}
\caption{Average identification accuracy as a function of the number of top-$L$ selected partitions for the four community-based fingerprint features: (a) $PC$, (b) $PC^+$, (c) $DC$, and (d) $DC^+$. Each panel shows results for all eight HCP tasks for sample size $S=100$, averaged over 10 independent runs. The vertical dashed line indicates the selected value $L=10$. Across all features, identification accuracy decreases as additional lower-ranked partitions are included, supporting the use of a small set of highly discriminative partitions in the proposed framework.}
\label{fig:SA_C}
\end{figure*}

To determine the appropriate number of partitions, we analyzed MPED as a function of the number of partitions across all tasks and fingerprinting metrics. We then identified the minimum number of rank-ordered partitions required to achieve 99\% of the maximum MPED value. This analysis consistently indicated that approximately the top-10 partitions were sufficient to capture nearly all of the discriminative information contained within the candidate partition set. In Fig.~\ref{fig:SA_S}, the top-10 partitions selected according to the 99th-percentile MPED criterion are highlighted in red. Each scatter plot combines partitions generated across 10 independent runs, with each run performed on a different randomly sampled subset of 100 subjects from the HCP dataset and evaluated independently. Most of these partitions are associated with high MPED values and high identification accuracy, further supporting the effectiveness of the proposed MPED-based partition selection strategy. To further validate the selection of top-$L$ partitions, we evaluated the identification accuracy as a function of the number of top-ranked partitions. As shown in Fig.~\ref{fig:SA_C}, the performance of the framework decreases as additional lower-ranked partitions are included. This finding suggests that lower-ranked partitions contribute limited discriminative information and supports the use of a small set of highly informative partitions.

\section{Sensitivity Analysis of Signed multilayer Modularity Parameters}
\label{Supp_sec1}
The  parameters $\gamma$ and $\omega$ play a critical role in determining the optimal joint community structure in SMM optimization. Rather than assuming the existence of a single optimal parameter pair, the proposed framework is based on the observation that different parameter combinations can uncover meaningful community organizations at distinct topological scales. Following \cite{betzel2017multi}, we generated partitions over a range of $\{\gamma,\omega\}$ values and discarded trivial or uninformative solutions, including singleton-community partitions, whole-network partitions, and partitions that exhibited identical community assignments across all layers. 

\begin{figure*}[t]
    \centering
    \includegraphics[width=\linewidth]{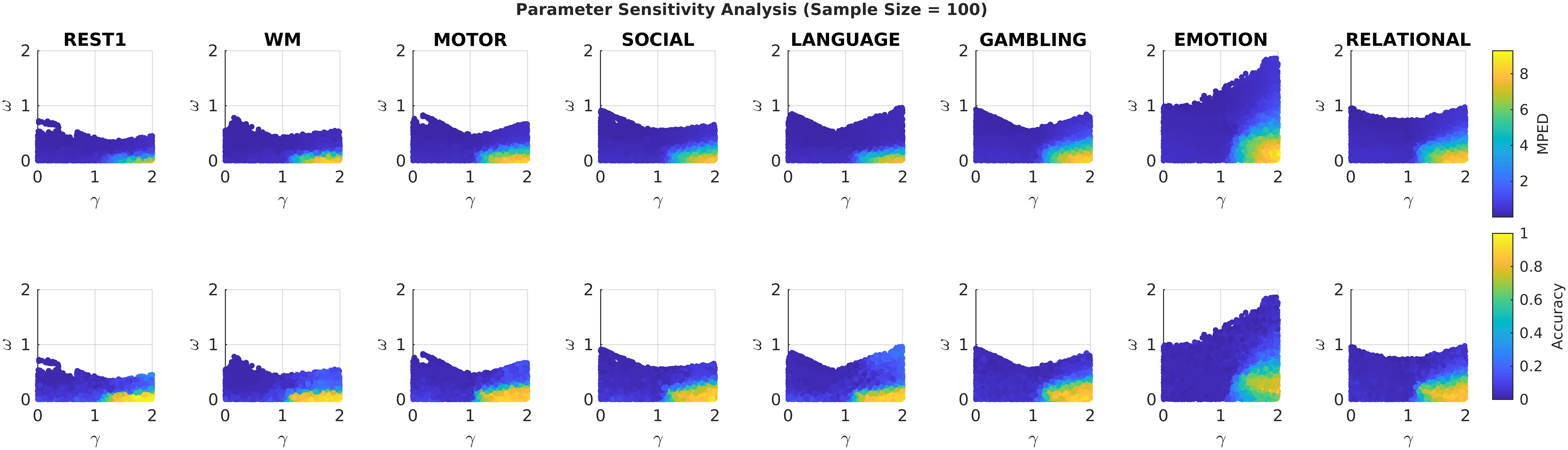}
    \caption{Sensitivity analysis of the signed multilayer modularity (SMM) parameters for all tasks ($S=100$). The top row shows the mean pairwise Euclidean distance (MPED), while the bottom row shows the corresponding identification accuracy across the $(\gamma,\omega)$ parameter space. Each point represents a single partition obtained from the parameter search. Results are aggregated across 10 independent runs, with each run generating 1,000 candidate partitions (10,000 partitions in total).}
    \label{fig:SA_SMM}
\end{figure*}

To assess the sensitivity of the proposed framework to parameter selection, we conducted a comprehensive analysis of the candidate partitions over the parameter ranges $\gamma \in [0,2]$ and $\omega \in [0,2]$, using both the unsupervised performance criterion, mean pairwise Euclidean distance computed from the extracted fingerprint vectors based on $PC^+$, and the supervised performance criterion, subject identification accuracy. The corresponding results for sample size \(S=100\) are presented in Fig.~\ref{fig:SA_SMM}. This figure shows both the MPED for the training data and subject identification accuracy for the test data based on the selected partition at different $(\gamma,\omega)$ pairs. Across tasks, high-performing partitions were found to be distributed over a region of the $(\gamma,\omega)$ parameter space rather than concentrated around a single parameter pair. This observation suggests that the proposed framework is robust to small variations in modularity parameters and does not rely on a unique optimal solution. Furthermore, the regions associated with high MPED values largely coincide with those yielding high identification accuracy, supporting the use of MPED as an effective criterion for selecting discriminative partitions.

\section{Discriminative Connectivity Maps (DCMs)}
\begin{figure*}[!t]
\centering
\captionsetup[subfigure]{labelformat=empty}

\begin{subfigure}[t]{0.24\textwidth}
\centering
\includegraphics[width=\linewidth]{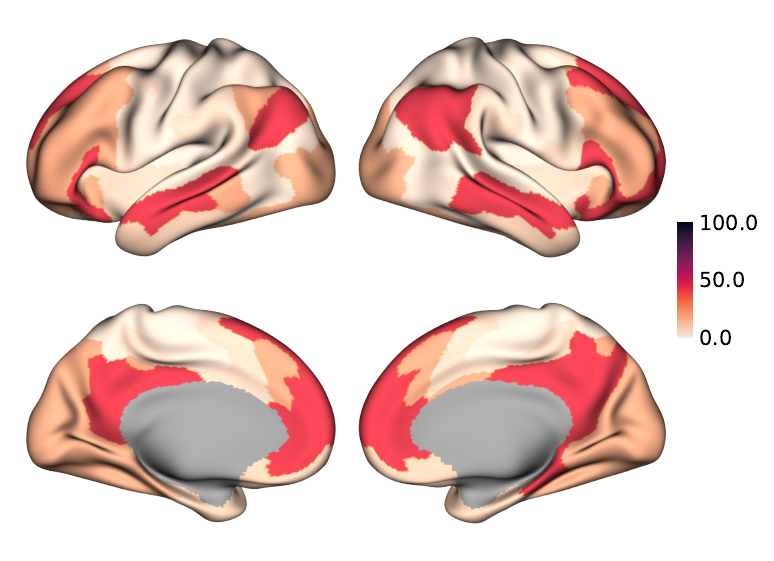}
\caption*{Rest-DCM}
\end{subfigure}
\begin{subfigure}[t]{0.24\textwidth}
\centering
\includegraphics[width=\linewidth]{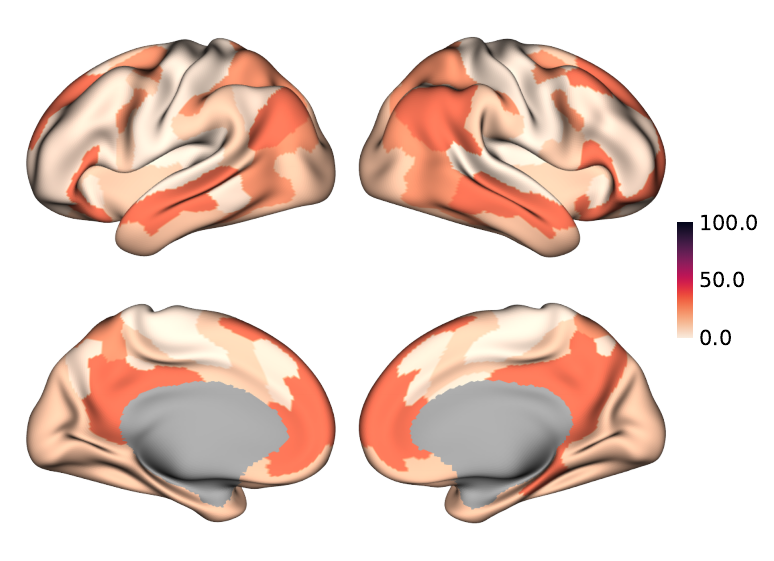}
\caption*{WM-DCM}
\end{subfigure}
\begin{subfigure}[t]{0.24\textwidth}
\centering
\includegraphics[width=\linewidth]{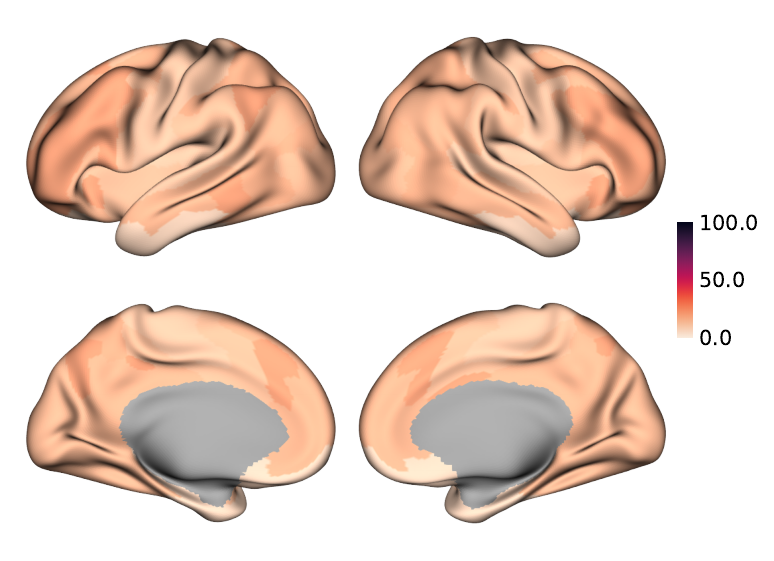}
\caption*{MOT-DCM}
\end{subfigure}
\begin{subfigure}[t]{0.24\textwidth}
\centering
\includegraphics[width=\linewidth]{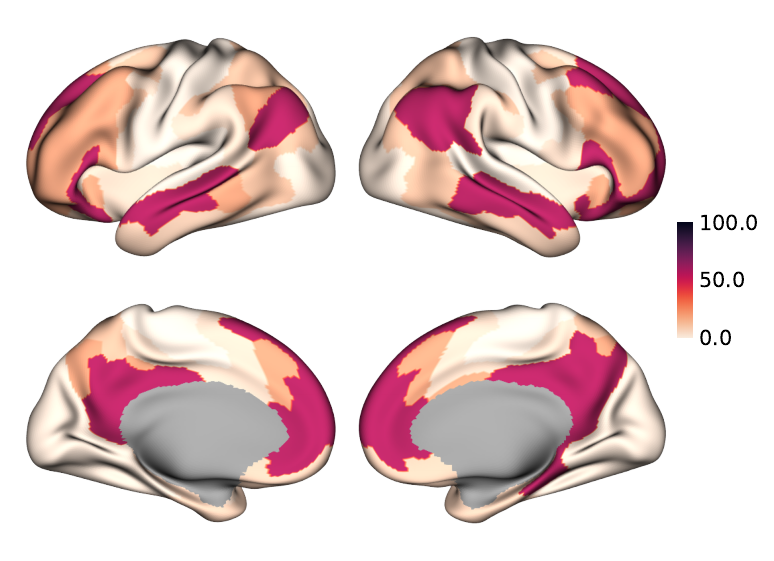}
\caption*{SOC-DCM}
\end{subfigure}

\vspace{0.2cm}

\begin{subfigure}[t]{0.24\textwidth}
\centering
\includegraphics[width=\linewidth]{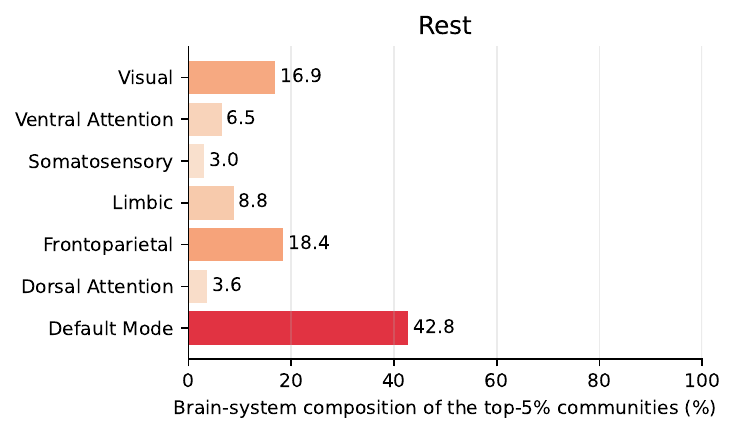}
\end{subfigure}
\begin{subfigure}[t]{0.24\textwidth}
\centering
\includegraphics[width=\linewidth]{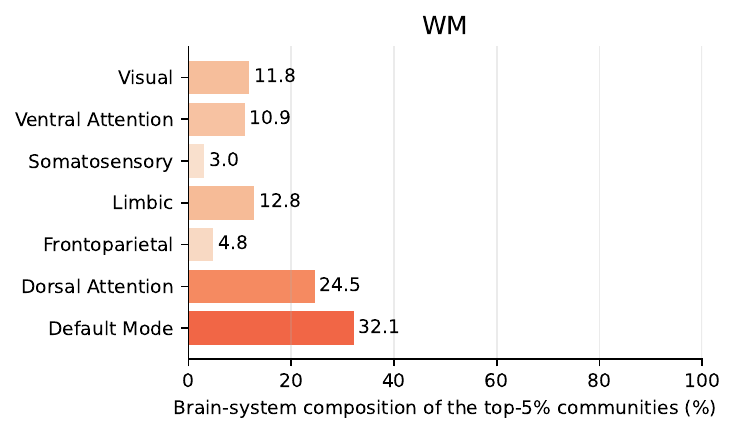}
\end{subfigure}
\begin{subfigure}[t]{0.24\textwidth}
\centering
\includegraphics[width=\linewidth]{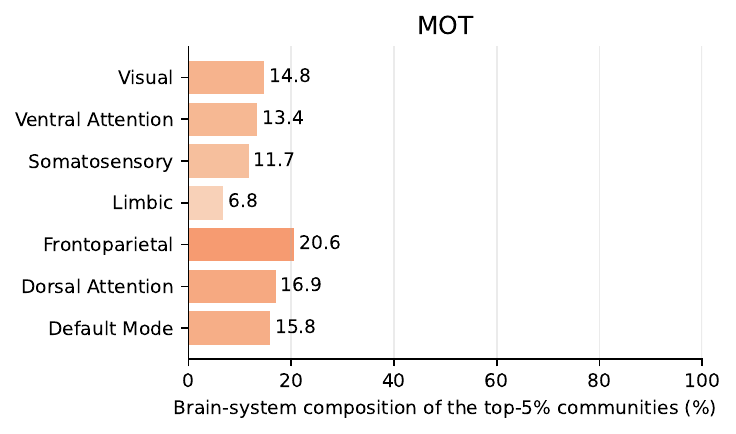}
\end{subfigure}
\begin{subfigure}[t]{0.24\textwidth}
\centering
\includegraphics[width=\linewidth]{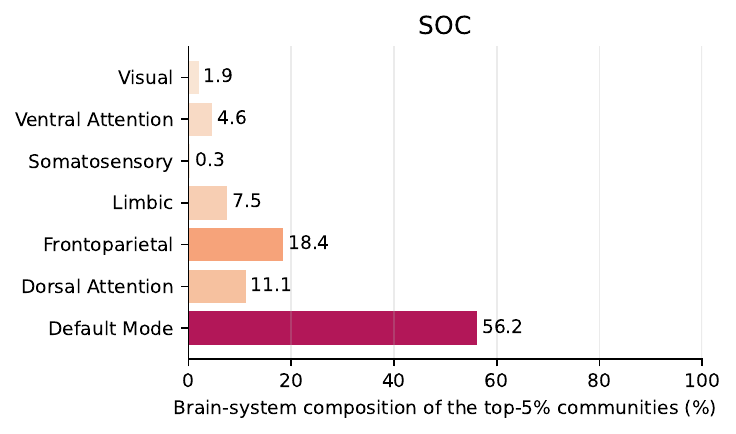}
\end{subfigure}

\vspace{0.5cm}

\begin{subfigure}[t]{0.24\textwidth}
\centering
\includegraphics[width=\linewidth]{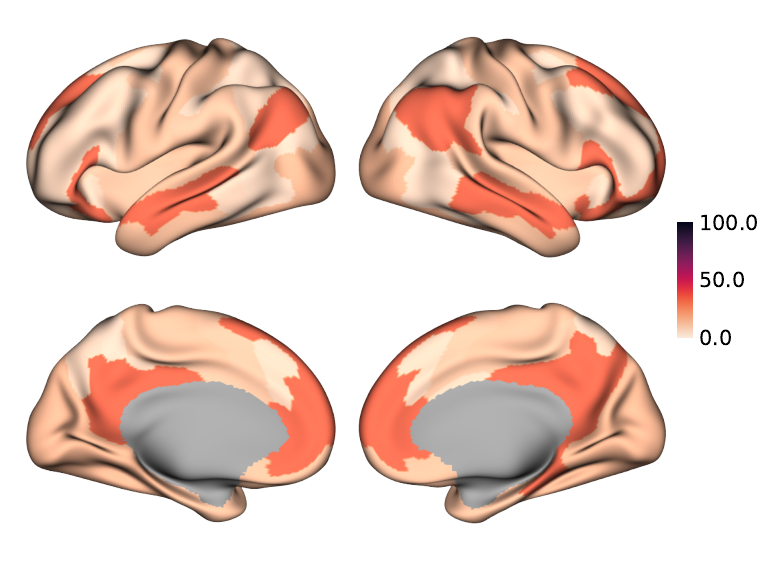}
\caption*{LAN-DCM}
\end{subfigure}
\begin{subfigure}[t]{0.24\textwidth}
\centering
\includegraphics[width=\linewidth]{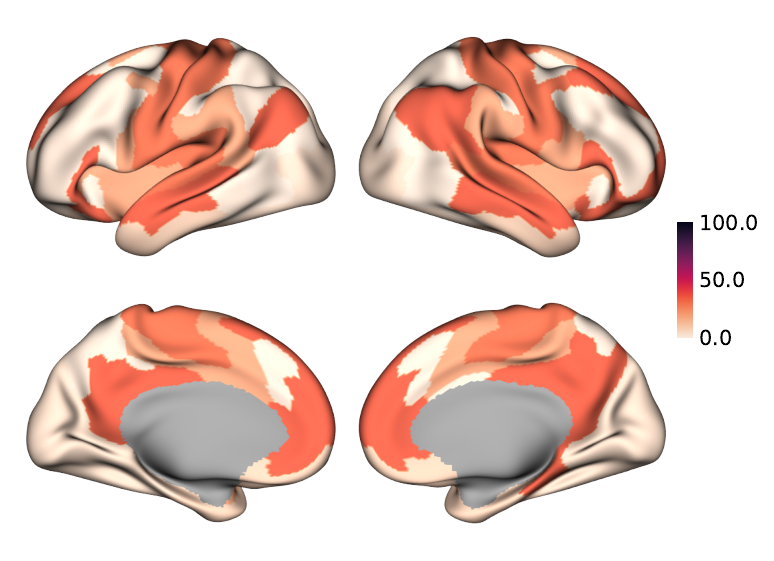}
\caption*{GAM-DCM}
\end{subfigure}
\begin{subfigure}[t]{0.24\textwidth}
\centering
\includegraphics[width=\linewidth]{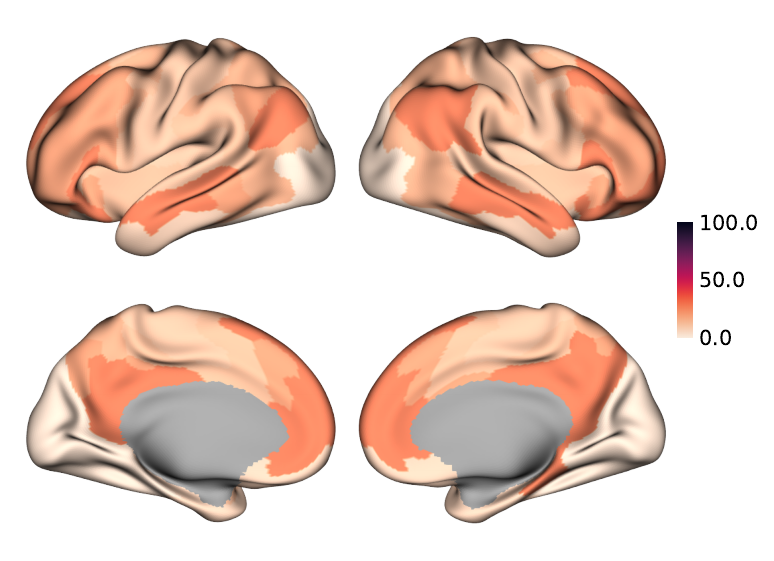}
\caption*{EMO-DCM}
\end{subfigure}
\begin{subfigure}[t]{0.24\textwidth}
\centering
\includegraphics[width=\linewidth]{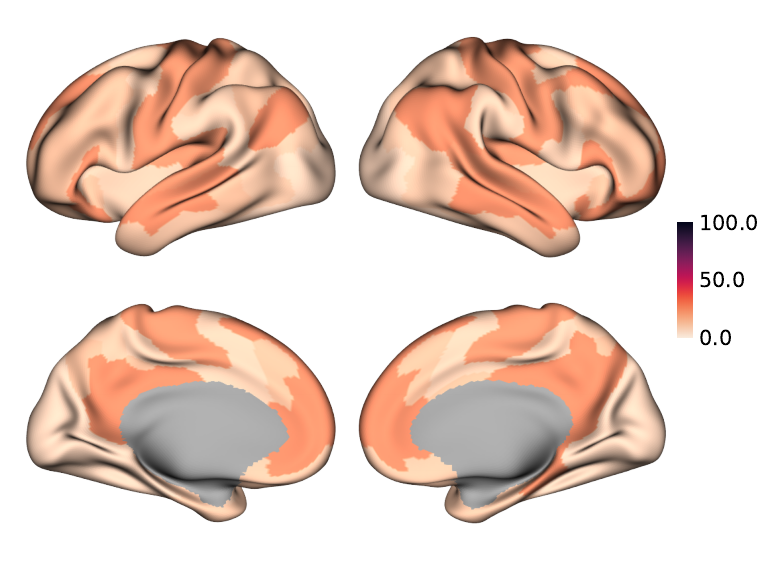}
\caption*{REL-DCM}
\end{subfigure}

\vspace{0.2cm}

\begin{subfigure}[t]{0.24\textwidth}
\centering
\includegraphics[width=\linewidth]{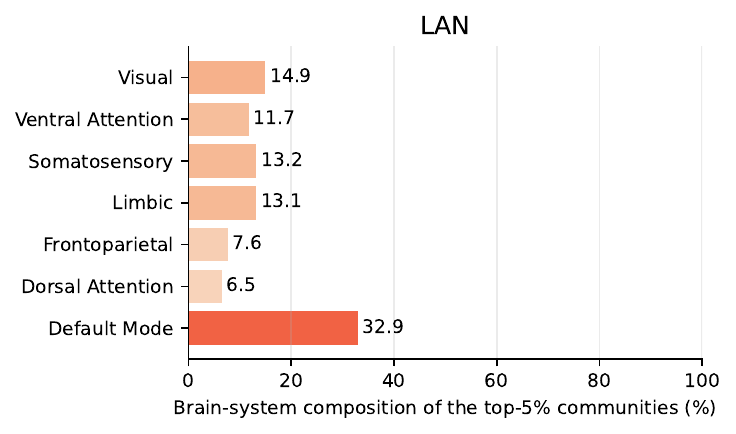}
\end{subfigure}
\begin{subfigure}[t]{0.24\textwidth}
\centering
\includegraphics[width=\linewidth]{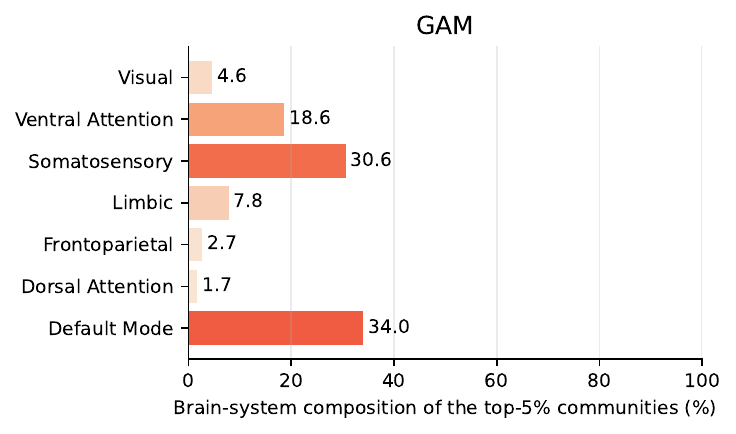}
\end{subfigure}
\begin{subfigure}[t]{0.24\textwidth}
\centering
\includegraphics[width=\linewidth]{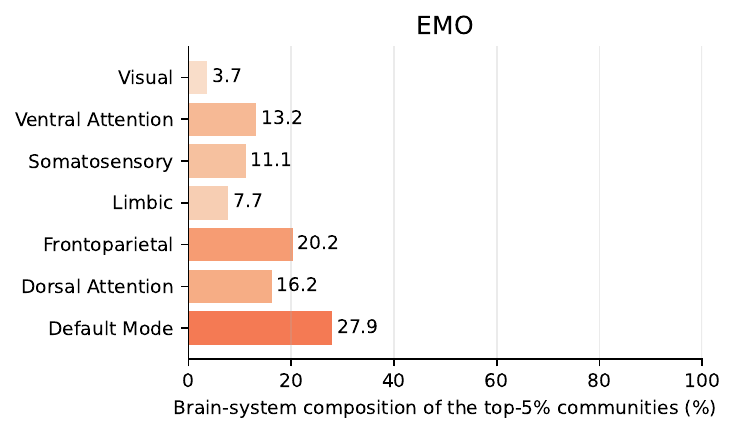}
\end{subfigure}
\begin{subfigure}[t]{0.24\textwidth}
\centering
\includegraphics[width=\linewidth]{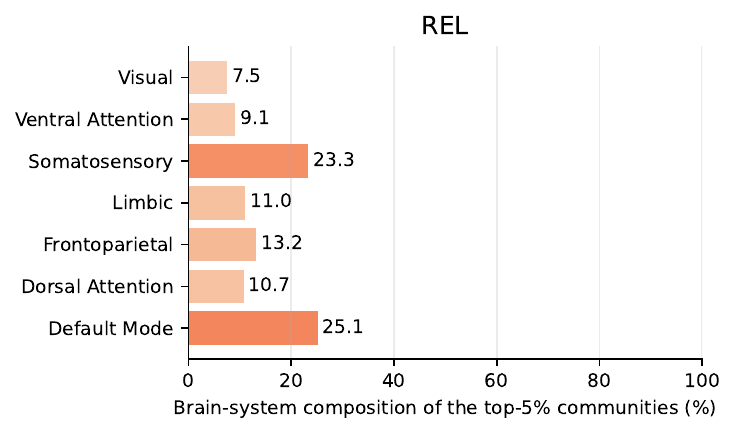}
\end{subfigure}
\caption{Discriminative connectivity maps (DCMs) illustrating the spatial distribution of the top-5\% discriminative community-based fingerprints (first and third rows), with the corresponding contributions of the seven Yeo functional brain systems (second and fourth rows) for each HCP task ($S=400$).}
\label{fig:Top5_400}
\end{figure*}

Fig.~\ref{fig:Top5_400} presents the DCMs obtained from the top-5\% most discriminative communities. Compared with the top-1\% DCMs (Fig. ~5 in the original paper), the overall spatial organization remains highly consistent across tasks, indicating that the discriminative connectivity patterns identified by the proposed framework are stable rather than driven by a small number of highly ranked communities. Although the contributions become more broadly distributed across multiple functional systems, each task retains its characteristic dominant network.

The resting-state condition remains primarily associated with the default mode network (DMN) (42.8\%), followed by the frontoparietal network (FPN) (18.4\%) and visual (16.9\%) systems, indicating that internally directed functional organization continues to dominate subject-specific fingerprints even after including additional discriminative communities. 

Similarly, the social task continues to exhibit the strongest contributions from the DMN (56.2\%), confirming that social cognition is consistently characterized by discriminative connectivity within association cortices involved in mentalizing and self-referential processing.

The working memory task preserves the prominence of the DMN (32.1\%) and the dorsal attention network (DAN) (24.5\%) systems, while the relative contribution of the FPN increases compared with the top-1\% analysis, reflecting a broader engagement of executive-control networks. 

The motor task remains distributed across FPN (20.6\%), DAN (16.9\%), DMN (15.8\%), visual (14.8\%), and somatosensory (SOM) (11.7\%) systems, suggesting that subject-specific motor fingerprints arise from coordinated interactions among multiple functional systems rather than isolated sensorimotor regions.

The language task continues to be dominated by the DMN (32.9\%), with substantial contributions from Limbic (13.1\%), SOM (13.2\%), ventral attention network (VAN) (11.7\%), and visual (14.9\%) systems, indicating a distributed organization supporting language-related individual variability. 

The gambling task exhibits comparable contributions from the DMN (34.0\%) and SOM (30.6\%) systems together with VAN (18.6\%), consistent with the integration of reward processing, sensory feedback, and attentional control. 

The emotion task remains dominated by the DMN (27.9\%) and FPN (20.2\%) systems, while relational reasoning continues to involve the DMN (25.1\%) and SOM (23.3\%) systems with additional contributions from FPN (13.2\%) and limbic (11.0\%) networks.

The top-5\% DCMs demonstrate that the spatial organization of discriminative connectivity remains task-specific while becoming more spatially distributed as additional high-ranking communities are incorporated.

\section{Between-Task Experiments}
\begin{figure*}[t]
    \centering
    \includegraphics[width=\textwidth]{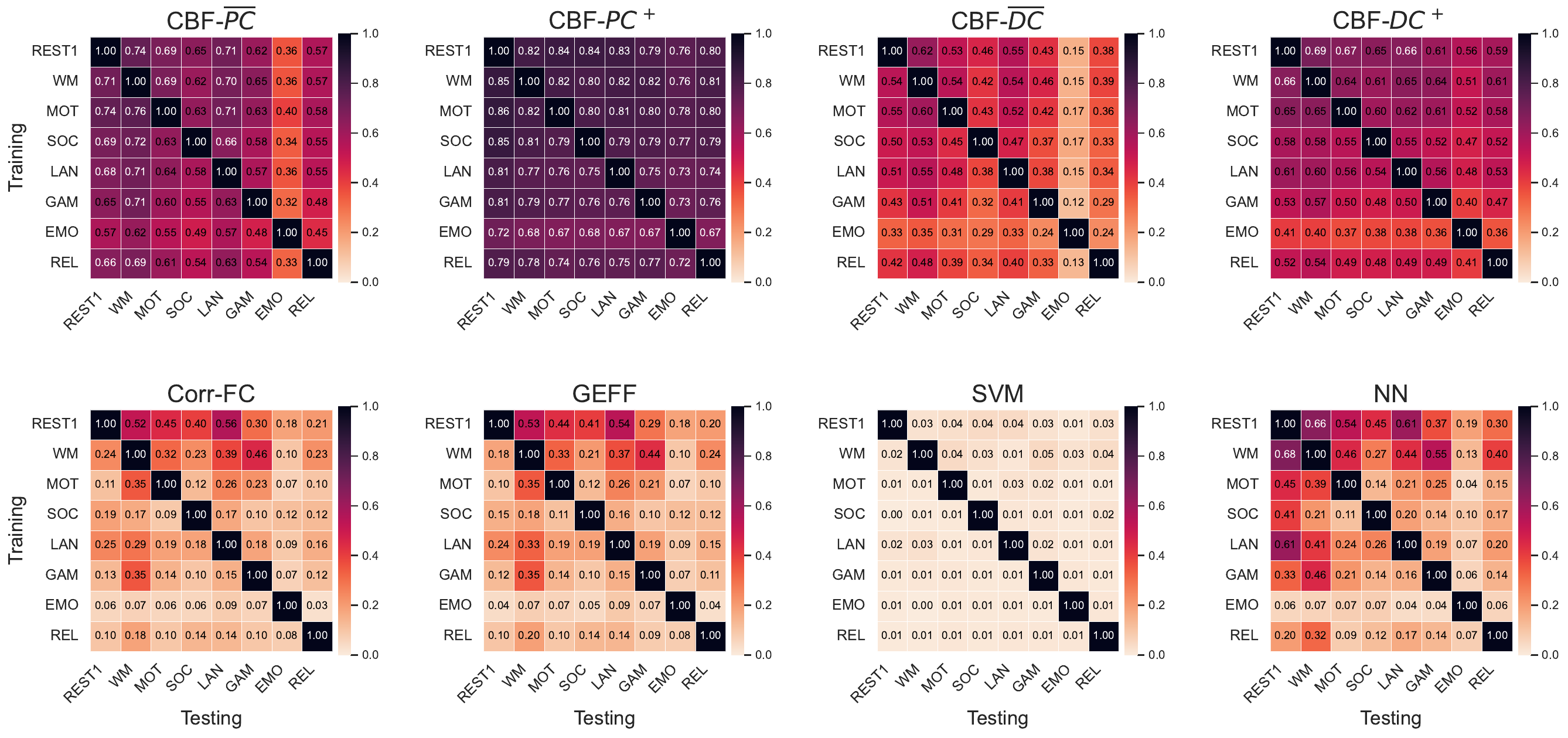}
    \caption{Between-task fingerprinting results across different methods for a sample size of $S=200$.} 
    \label{fig:between_200}
\end{figure*}

\begin{figure*}[t]
    \centering
    \includegraphics[width=\textwidth]{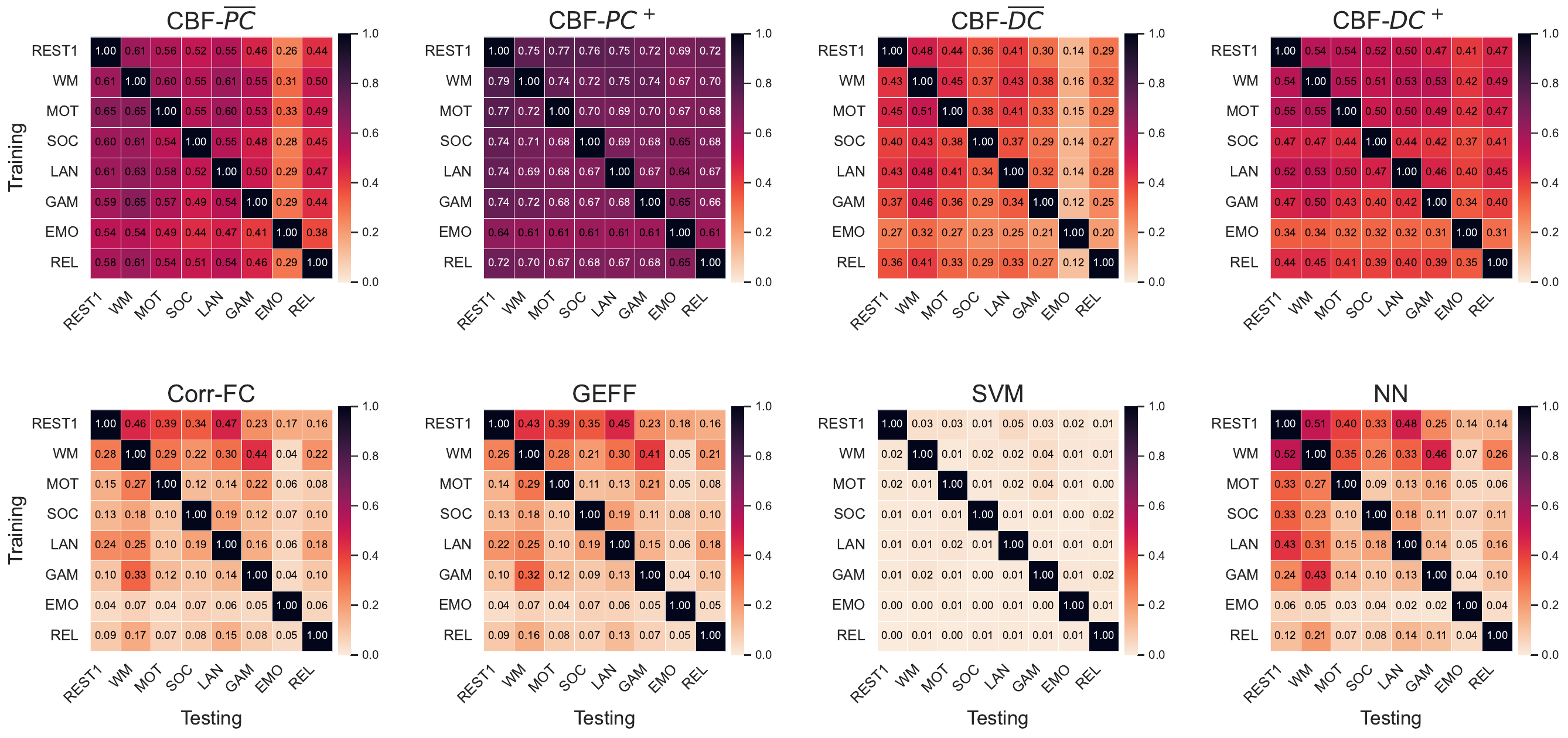}
    \caption{Between-task fingerprinting results across different methods for a sample size of $S=300$.} 
    \label{fig:between_300}
\end{figure*}

Between-task fingerprinting results across different methods for sample sizes of $S=200$, and $300$ are presented in Figs. \ref{fig:between_200} and \ref{fig:between_300}, respectively.

\bibliographystyle{unsrt}
\bibliography{refs-sema}

